\documentclass[fleqn,usenatbib]{mnras}

\usepackage{newtxtext,newtxmath}
\usepackage[T1]{fontenc}
\usepackage{tabularx}

\DeclareRobustCommand{\VAN}[3]{#2}
\let\VANthebibliography\thebibliography
\def\thebibliography{\DeclareRobustCommand{\VAN}[3]{##3}\VANthebibliography}

\usepackage{graphicx}	
\usepackage{amsmath}	
\usepackage{multicol}

\usepackage[dvipsnames]{xcolor} 

\title[Evaluating CHE in stellar binaries]{Evaluating chemically homogeneous evolution in stellar binaries: Electromagnetic implications - Ionizing photons, SLSN-I, GRB, Ic-BL}

\author[S. Ghodla et al.]{Sohan Ghodla$^{\thanks{sgo069@aucklanduni.ac.nz}{1}}$,  J. J. Eldridge$^{1}$,  Elizabeth R. Stanway$^{2}$ , Héloïse F. Stevance$^{1, 3}$ \\
$^{1}$Department of Physics, University of Auckland, Private Bag 92019, Auckland, New Zealand \\ $^{2}$Department of Physics, University of Warwick, Gibbet Hill Road, Coventry, CV4 7AL, UK \\ $^{3}$Astrophysics Research Centre, School of Mathematics and Physics, Queen’s University Belfast, N. Ireland, BT7 1NN, United Kingdom
}

\date{Accepted XXX. Received YYY; in original form ZZZ}

\pubyear{2022}

\begin{document}
\label{firstpage}
\pagerange{\pageref{firstpage}--\pageref{lastpage}}
\maketitle


\begin{abstract}

We investigate the occurrence of rapid-rotation induced chemically homogeneous evolution (CHE) due to strong tides and mass accretion in binaries.
To this end, we generalize the relation in Packet (1981) to calculate the minimum angular momentum (AM) accretion required by a secondary star to experience accretion-induced CHE.
Contrary to traditionally assumed 5-10 per cent accretion of initial mass ($Z \lesssim 0.004$, $M  \gtrsim$ 20 M$_{\odot}$) for spinning up the accretor (resulting in CHE) this value can drop to $\sim$ 2 per cent for efficient AM accretion while for certain systems it could be substantially larger.
We conduct a population study using \textsc{bpass} by evolving stars under the influence of strong tides in short-period binaries and also account for the updated effect of accretion-induced spin-up.
We find accretion CHE (compared to tidal CHE) to be the dominant means of producing homogeneous stars even at 10 per cent AM accretion efficiency during mass transfer. Unlike tidal CHE, it is seen that CH stars arising due to accretion can retain a larger fraction of their AM till core collapse.
Thus we show that accretion CHE could be an important formation channel for energetic electromagnetic transients like GRBs, Ic-BL (SLSN-I, Ic-BL) under the collapsar (magnetar) formalism and a single CH star could lead to both the transients under their respective formation scenario. Lastly, we show that under the current treatment of CHE, the emission rate of ionizing photons by such stars decreases more rapidly at higher metallicities than previously predicted.

\end{abstract}




\section{Introduction} \label{sec: intro}

Along with mass and metallicity, rotation is considered a crucial initial parameter in shaping the evolution of massive stars \citep{Maeder2000, Heger2000, Langer_2012_review_massive_stars}.
The effect of rotation can vary, but one scenario where it can significantly alter stellar evolution is during the occurrence of quasi-chemically homogeneous evolution (CHE). \cite{Maeder1987_CHE} showed that as a massive star evolves through its main sequence (MS), for a sufficiently rapid rotation rate, internal mixing can occur at a pace faster than the growth of the chemical gradient. 
Over time this could altogether impede the formation of a core-envelope structure and the system instead evolves into a quasi-chemically homogeneous helium star.
Hence, unlike the structural profile possessed by the slow rotators which subsequently leads to their expansion into red supergiants, CH stars on the other hand remain compact and nearly fully mixed (e.g., \citealt{Yoon_2005, Yoon_Langer2006_Xrbs, Woosley_Heger2006-Xrbs, Cantiello2007, deMink2009, deMink_Mandel2016, Song_2016, Marchant2016, Aguilera-Dena_2018, du_Buisson2020, Riley_2021}).
Moreover, such stars can more easily retain their  spin angular momentum (AM) as compared to slow rotators.
This is because the lack of a significantly large hydrogen-rich envelope, in particular during the later stages of CH stars, implies that magnetic torque induced spin down due to the Spruit-Tayler dynamo \citep{Spruit2002} is minimal (e.g., \citealt{Maeder_Meyent_2003, Petrovic_2005, Heger_2005, Yoon_Langer2006_Xrbs}).



\begin{figure}
 \centering
  \vspace{-0.75cm}
 \includegraphics[width = 1.02\linewidth]{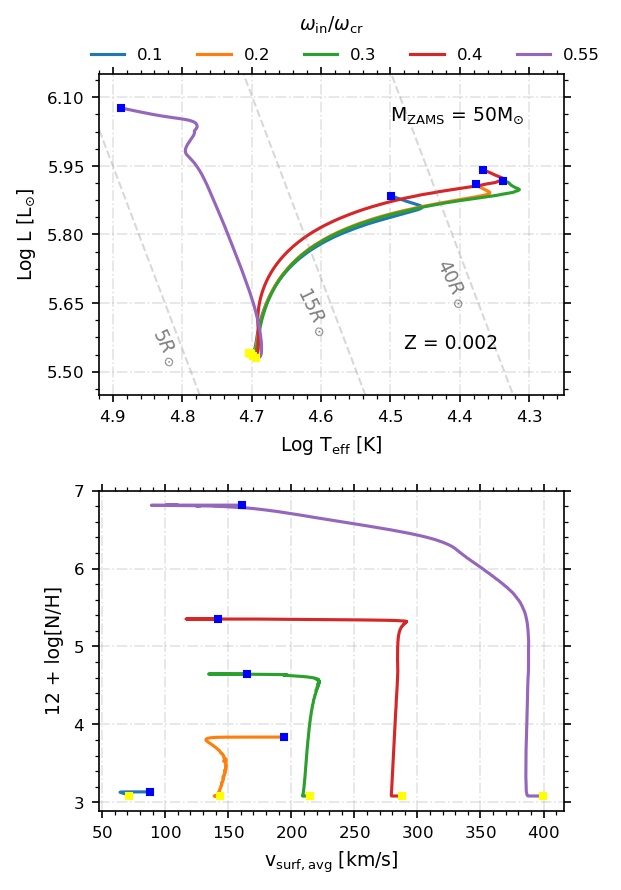}
 \vspace{-0.3cm}
 
 \caption{The top figure shows the HR diagram of a 50 M$_{\odot}$ star with initial metallicity $Z = 0.002$ for a range on initial angular velocities ($\omega_{\rm in}$) normalised by the critical angular velocity ($\omega_{\rm cr}$). Most noticeably the tracks experience a bifurcation for sufficiently rapid rotation. This is the trait of a CH star. The bottom plot shows the surface nitrogen mass fraction for these stars. The yellow squares are the points when the star is on zero-age MS while the blue is when core hydrogen has exhausted.}
 \label{fig: HR_diagraml}
\end{figure}

As an example, Fig. \ref{fig: HR_diagraml} demonstrates some of the  presently accepted observational traits of rotating stars. Slowly rotating stars experience typical stellar evolution with stars becoming cooler on the MS as they become more luminous.
In comparison a sufficiently rapid rotation rate causes the tracks on the Hertzsprung-Russel diagram to bifurcate. This signals the onset of CHE causing the star to now appear more luminous and bluer.
Rotation also leads to the internal transport of material and consequently brings processed matter, most notably nitrogen (see Fig.  \ref{fig: HR_diagraml}), from the convective core towards the radiative envelope.

Recent observations and theoretical models hint that rapidly rotating massive single stars might not be very numerous in nature. 
In particular, from the observation of the projected rotational velocity of O-type stars in LMC 30 Doradus region, \cite{Ramirez_Agudelo_2013_rotation_evidence} concluded that most of the rapid rotators were either present in binaries or were a merger product of a binary system.
Further evidence comes from the analysis of the O star population of six nearby Galactic open stellar clusters by \cite{Sana2012} who found that a majority of massive stars were in binaries with about two-thirds of them able to interact during their lifetime.
Motivated by this, \cite{deMink2013} using stellar models concluded that mass transfer and binary mergers could be the main source of rapid rotation in massive stars. 


As such, in the light of recent results, binary interaction seems to be the dominant channel of generating rapid rotators. Consequently, here we discuss two pathways of inducing rapid stellar rotations that could lead to CHE for stars in binaries. 
These are applicable to stars present in (i) short period massive binaries where tides are strong enough such that the rotational frequency of the star becomes synchronous with the orbital frequency also known as tidal CHE (see \citealt{deMink2009, Song_2016}) and (ii) systems with weak tidal interactions where a sufficient amount of accretion can induce CHE in the secondary component of the binary (e.g. \citealt{Cantiello2007}). 
This is referred as accretion CHE here. Both scenarios prefer a low metallicity environment as due to weak stellar winds on the MS, these stars experience a lesser AM-loss induced spin down (e.g. see, \citealt{Yoon_Langer2006_Xrbs, Woosley_Heger2006-Xrbs}). 
In the past, tidal CHE has been studied by \cite{deMink2009, Song_2016, deMink_Mandel2016, Marchant2016, Qin_2019, du_Buisson2020, Riley_2021} and others.
On the other hand, accretion CHE was initially introduced by \cite{Cantiello2007} to provide a viable mechanism for generating long Gamma-Ray Bursts (LGRBs). Further analysis in the context of population synthesis studies has been primarily conducted by \cite{Eldridge_2011, Bpass2017, Bpass_2018, Chrimes_2020}.

A number of LGRBs have been simultaneously observed with  core-collapse supernova events (e.g. \citealt{Galama_1998, Stanek_2003, Sollerman_2006}). As such these rapid rotators have been thought of as potential progenitors of the observed LGRBs under the collapsar scenario \citep{Woosley1993_collapsar}. 
CH stars have recently been proposed as an additional pathway for forming short-period massive binary black holes \citep{deMink_Mandel2016, Marchant2016} with a possibility of small merger delay time. They could also be the source of the spin of a subsequently formed neutron star or black hole. 
Additionally, CH stars have been used in the past to fit the observed spectra of ionizing photons from high redshift galaxies \citep{Eldridge_Stanway_2012, Szecsi_2015,  Stanway2016, Xiao_2018, Sibony_2022}. Observationally, \cite{Bouret2003} speculated that CHE might be the cause of the high nitrogen surface abundance of a young, hot star (NGC 346-355) present in the SMC. 
Similarly, \cite{Martins2009_CHE_obs} analyzed a Wolf-Rayet star in the SMC, \cite{Koenigsberer_2014_Obs_CHE} a Luminous-Blue-Variable/Wolf-Rayet binary (HD 5980) in the SMC, and \cite{Shenar2017}  the Wolf-Rayet binary R145 in the LMC and each concluded that the systems experienced CHE in the past. 
Additionally, \cite{Almeida_2015_Obs_CHE} discovered a massive over-contact O-type binary, VFTS 352 in the 30 Doradus region where both the stars are too hot for their respective masses and could only explain their characteristic invoking enhanced mixing due to CHE.
\cite{Martins2013_Obs_CHE_solarZ} have also used CHE to explain the surface properties and HR diagram location of some H-rich early-type WN stars in the LMC and our Galaxy. Surprisingly, they report that these stars could have a metallicity ($Z$) ranging from 0.6Z$_{\odot} < Z < $ Z$_{\odot}$ (Z$_{\odot} = 0.02)$. 
Though many of the above observations rely on rapid rotation to explain the surface abundance anomalies and higher temperature and luminosity, \cite{Hunter_2008_surface_anomalies} found two distinct groups of stars that are either rapid rotators with no surface anomalies or have surface anomalies but are not rapidly rotating. Subsequent studies show that it is possible that the surface anomalies are primarily caused by binary interactions, e.g. \cite{deMink2013}. Consequently, the applicability of CHE as a possible explanation for the characteristics of the observed systems in the LMC/SMC remains ambiguous.

\textsc{compas} \citep{COMPAS_2022}, a rapid stellar population synthesis code, has integrated the tidal CHE pathway such that it could simultaneously explore its interaction with the rest of the conventionally evolved stellar population within the framework of a single code. Following in the footsteps of \cite{Riley_2021}, here we implement tidal CHE in \textsc{bpass} (Binary population and spectral synthesis suite), a detailed stellar population synthesis code. Secondly, we improve the existing prescription for accretion CHE. We generalize the analytical relation in \cite{Packet1981} to give the threshold mass accretion to spin up a star to any desired (below critical) angular velocity. We then use \textsc{mesa} stellar evolution models in conjunction with the above-derived relation to calculate the threshold mass accretion required by a star (as a function of its initial mass and metallicity) to spin up just enough to experience CHE. These additions allow us to compare the population of CH stars arising from both pathways.

The aim of this article is to improve the CHE implementation within population synthesis studies and evaluate which among accretion or tides is the dominant source for producing such systems. We then investigate some of the electromagnetic implications of these systems including potential LGRB, broadlined type-Ic (Ic-BL) SN, superluminous SN type-I (SNSN-I) and ionizing photon sources. We show that under suitable conditions a CH star could produce both GRB and Ic-BL (SLSN-I and Ic-BL) under the collapsar (magnetar) scenario. In a subsequent article, we would look into the implication of the CHE pathway for the gravitational wave transients and cosmic rates. The remainder of the article is organized as follows: in section \ref{sec: Methods} we outline the \textsc{mesa} models and discuss the minimum threshold initial angular velocity for CHE followed by the minimum threshold accretion for CHE in section \ref{sec: threshold_accretion_CHE}. Then section \ref{sec: CHE in BPASS} discusses the implementation of CHE utilising the \textsc{bpass} suite. We present some observational implications of CHE in section \ref{sec: Results} followed by discussion and some concluding remarks in section \ref{sec: Discussion} and \ref{sec: conclusion}.

\vspace{-0.5cm}
\section{The threshold initial angular velocity for CHE from MESA models} \label{sec: Methods}

Rotation affects the hydrostatic and thermal equilibrium of a star, eventually impacting the stellar structure.
According to \cite{Zeipel1924}, for rotating stars, the equation of radiative transport takes the form, ${\rm flux} \propto {g}_{\rm eff}$ implying that $T(\psi) \propto{ {g}_{\rm eff}^{1/4}(\psi)}$, where $\psi$ are the equipotential surfaces, $T$ is temperature and ${g}_{\rm eff}$ the effective gravitational acceleration such that $\nabla \psi = {g}_{\rm eff}$. Since centrifugal force leads to a non-uniform value of $g_{\rm eff}(\psi)$, this implies that there should be higher radiative flux at the poles than at the equator (leading to gravity darkening). 

Consequently, the equipotential surfaces would not simultaneously be surfaces of constant heat flux, hence disturbing the state of hydrostatic equilibrium.
This subsequently leads to large-scale meridional circulation, rising at the poles and descending at the equator \citep{Eddington1925, sweet1950}.
These meridional currents along with convection and turbulence diffusion (induced due to various instabilities, see Sec.\ref{MESA_outline}) aid the transport of AM \footnote{As mentioned in section \ref{sec: intro}, AM is also transported due to Spruit-Tayler dynamo.} and chemical elements within the star. 
The vertical transport of AM and chemical species is directly proportional to the vertical velocity of the meridional currents, the angular velocity of the star $\omega$, and the diffusion coefficient.
Moreover the chemical mixing timescale ($\tau_{\rm mix})$ and more importantly the Eddington-Sweet timescale ($\tau_{\rm ES}$) are inversely proportional to $\omega$  and mass ($M$) such that with increasing $M$ and $\omega$ both $\tau_{\rm mix}$ and $\tau_{\rm ES}$ scale down at a faster rate (and becomes smaller) than the MS lifetime of the star (e.g. \citealt{Maeder2000}). As such massive rapid rotators are favorable systems for efficient chemical mixing.

\begin{figure}
    \centering
     \vspace{-0.1cm}
    \includegraphics[width = 1.02\linewidth]{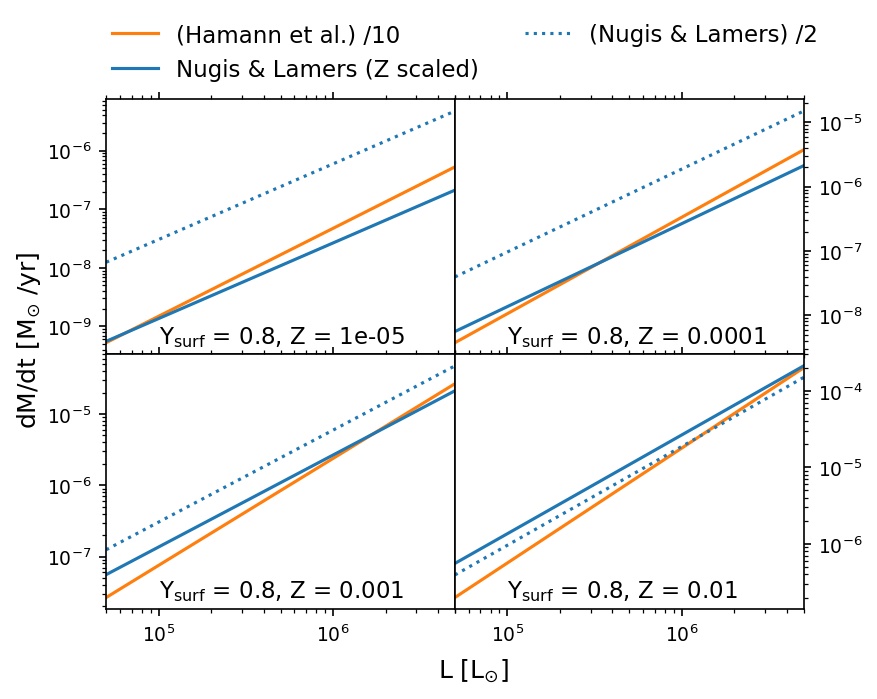}
    \vspace{-0.5cm}
    \caption{A comparison between the initial metallicity scaled \protect\cite{Nugis_Lamers_2000} WR wind scheme (blue) to the widely used \protect\cite{Hamann_1995} relation down-scaled by a factor of 10 (e.g. \protect\citealt{Yoon_2005, Brott2011}). The surface mass fraction of helium $Y_{\rm surf}$ is taken as 0.8. The \protect\cite{Nugis_Lamers_2000} wind scheme reduced by half (Dutch scaling factor = 0.5) is also shown.}
    \label{fig:Mdot_comparison}
\end{figure} 

\subsection{MESA models outline} \label{MESA_outline}
To make rotating stellar models for investigating CHE, we use the 1D stellar evolution code \textsc{mesa} r15140 \citep{MESA2011, MESA2013, MESA2015, MESA2018, MESA2019}. \textsc{mesa} treats rotation in the form of shellular approximation, i.e. the angular velocity is assumed constant over an isobaric surface \citep{Zahn1992_Shellular_rotation, Meynet_Maeder_1997_shellular_rot}.  Our goal is to calculate the threshold minimum initial angular velocity $\omega_{\rm in}$ as a function of the initial mass and metallicity for an isolated star to experience CHE.
We choose initial mass $\in$ [15-300] M$_{\odot}$ (separated by 10 M$_{\odot}$ till 100 M$_{\odot}$ and by 25 M$_{\odot}$ subsequently) and initial metallicity mass fraction  $Z \in [10^{-5}, 10^{-4}, 3 \times 10^{-4}, 6 \times 10^{-4}, 0.001, 0.002, 0.003, 0.004, 0.005]$ and evolve them till core carbon depletion (as discussed later). The value of solar metallicity is taken as $Z_{\odot} = 0.02$ to be consistent with the earlier \textsc{bpass} framework.

There exist free parameters whose magnitudes are not determined by first principles but are rather extracted by matching numerical stellar models with observations.
As such to model convective energy transport, here we adopt the  mixing length theory with free parameter $\alpha_{\rm MLT} = 1.5$.
Convective mixing is treated as a step decay process with overshoot parameter $l_{\rm ov} = 0.185H_{P}$ where $H_{P}$ is the local pressure scale height.
These values have been obtained by calibrating the \textsc{mesa} models to existing \textsc{bpass} models in the 30-80 M$_{\odot}$ range. Semi convection is treated as in  \cite{Langer1985_Semiconvection} with the efficiency parameter $\alpha_{\rm SEM} = 1$ as in \cite{Langer1991_Semiconv} and  we use the Ledoux criterion to test for stability against convection.  

For wind mass loss we follow the Dutch wind mass loss prescription as a combination of \cite{deJager_1988, Vink_2001} and \cite{Nugis_Lamers_2000}. We follow \cite{Vink_mdot_WR_2005} - who found WR mass loss is metallicity dependent - and scale the \cite{Nugis_Lamers_2000} WR mass-loss rates rate with the initial stellar metallicity $Z$ \citep{Eldridge_Vink_2006} using the factor $(Z/Z_{\odot})^{0.5}$ as shown below

\begin{equation}
\dot{M} \simeq (Z/Z_{\odot})^{0.5} \cdot [10^{-11}\left(L / L_{\odot}\right)^{1.29} Y^{1.7} \textbf{Z}^{0.5}]
\label{eq: mdot_scaling}
\end{equation} 

\noindent where $\dot{M}, L, Y, \textbf{Z}$ represent the time-dependent mass loss rate, luminosity, helium mass fraction and metallicity mass fraction of the star respectively (we note that in Eq. \ref{eq: mdot_scaling}, $Z$ represents initial metallicity while $\textbf{Z}$ is time dependent). The above initial metallicity scaling allows accommodating the observed decrease in the relative population of WC over WN stars at lower $Z$ \citep{Eldridge_Vink_2006, Bpass2017}. This scaling factor then replaces the usually used Dutch scaling factor typically taken as a constant. The effect of our assumption of the mass-loss rate is shown in Fig. \ref{fig:Mdot_comparison}. We also include the rotation induced mass loss as $ \dot{M}(\omega)=\dot{M}(0) (1- \omega / \omega_{\text {crit }})^{- \xi}$ with $\xi = 0.43$ \citep{Langer1998}. As angular velocity $\omega$ approaches its critical value $\omega_{\rm crit}$,  this expression begins to diverge hence we set the variable \texttt{implicit\_mdot\_boost} = 0.1 in \textsc{mesa} to artificially boost the mass loss rate until the rotation falls below critical.


We consider various instabilities caused by rotation such as secular and dynamical shear instability, Eddington-Sweet circulation and Goldreich Schubert-Fricke instability. Following \cite{Yoon_Langer2006_Xrbs} we set the parameter $f_{\mu} = 0.1$ to account for the inhibiting effect of chemical gradients on the efficiency of rotational mixing.
They found that choosing this value offered a good match to observed surface abundances of helium and nitrogen. The efficiency of AM mixing is regulated by setting $f_{\rm c} = 1/30$ following \cite{Zahn_1992}. 

\begin{figure}
    \centering
     \vspace{-0.6cm}
    \includegraphics[width = 1\linewidth]{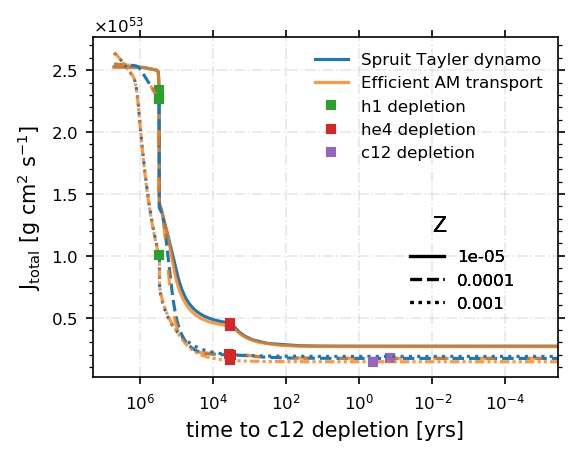}
    \vspace{-0.5cm}
    \caption{A comparison between the Spruit-Tayler (ST) dynamo induced AM transport \protect\cite{Spruit2002} and the efficient AM transport as described in \protect\cite{Fuller_2019} for a 60 M$_{\odot}$ CH star at three different metallicities. The blue and orange lines depict the ST and efficient AM transport mechanism respectively. The various linestyles represent the different metallicities and the green, red and purple colored boxes mark core hydrogen, helium and carbon depletion respectively. Efficient AM transport does not make any noticeable deviation in the total AM ($J_{\rm total}$)  evolution of the star when compared with ST dynamo. The stellar model with $Z = 10^{-5}$ fails before reaching core carbon depletion. Nonetheless, there is no significant AM loss post core helium depletion.}
    \label{fig: ST_vs_efficient_AM}
\end{figure}

\cite{Yoon_2005} have shown that for systems capable of experiencing CHE, magnetic torques are inefficient in transporting AM from the core to the surface as a core-envelope type of configuration is avoided. Nevertheless, we do implement the Spruit-Tayler (ST) dynamo \citep{Spruit2002} in our models. Recent asteroseismic observations reveal that cores of evolved stars rotate slower than predicted (e.g. see \citealt{Eggenberger_2019, Fuller_2019} and reference therein).
Motivated by this \cite{Fuller_2019} proposed that efficient transport of AM in the stellar interior is required to slow down their spin. Fig. \ref{fig: ST_vs_efficient_AM} compares the traditional ST dynamo induced AM transport \citep{Spruit2002} with the efficient AM transport \citep{Fuller_2019} for a 60 M$_{\odot}$ CH star, evolved till core carbon depletion at three different metallicities.
Unsurprisingly, the evolution of the total AM ($J_{\rm total}$) under the assumption of efficient AM transport closely matches that via the ST dynamo (also see \citealt{Fuller_Ma_2019}).
This is again due to the non-presence of a core-envelope bifurcation in CH stars which limits the magnetic torque-induced AM loss. As such in our current study, we do not employ the proposed efficient AM transport mechanism. Fig. \ref{fig: ST_vs_efficient_AM} is further discussed in section \ref{sec: Results}.

\begin{figure}
    \centering
     \vspace{-0.3cm}
    \includegraphics[width = 1\linewidth]{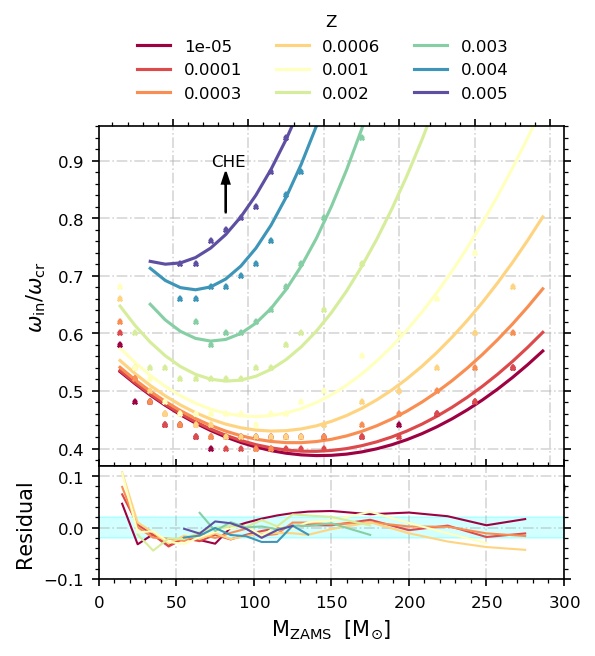}
    \vspace{-0.5cm}
    \caption{The threshold minimum initial angular velocity $\omega_{\rm in}$ as a function of initial mass for the isolated star ($M_{\rm ZAMS}$) to experience CHE. The legend represents the initial metallicity of the star. The cutoff metallicity lies somewhere between $Z = (0.005-0.006)$. The values of y-axis have been normalized with the critical initial angular velocity of the star assuming initial solid body rotation. Typically the fits have a residual error of $\sim 0.02$ as shown in the bottom subplot (cyan strip). The lower mass limit for CHE lies near $\sim 15$  M$_{\odot}$ ($\sim 30$  M$_{\odot}$ for $Z \geq 0.003$) and the upper limit is set when $\omega_{\rm CHE}/\omega_{\rm cr}$ approaches 1}
    \label{fig: Threhold_omega_Dutch_factor_variable}
\end{figure}

\subsection{The minimum threshold initial angular velocity for CHE}

It is seen (e.g. \citealt{Riley_2021}) that  at the end of the main sequence, stars for which the difference between surface and central hydrogen mass fraction $\Delta X < 0.2$ evolve in a CH manner. Hence we use this property as an indicator of CHE \footnote{Some models at $Z \geq 0.003$ that marginally fail to meet this criterion but still show the characteristics of a CH star are also included by extrapolating the fit in Fig. \ref{fig: Threhold_omega_Dutch_factor_variable} to $M_{\rm ZAMS}$ = 30 M$_{\odot}$.}.
Fig. \ref{fig: Threhold_omega_Dutch_factor_variable} shows for the \textsc{mesa} models the minimum initial angular velocity $\omega_{\rm in}$ for an isolated star of given $M_{\rm ZAMS}$ to experience CHE. The resulting threshold $\omega_{\rm in}$ is referred as $\omega_{\rm CHE}$. We fit the data to a parabolic trend to acquire $\omega_{\rm CHE} /\omega_{\rm cr}$ as a function of $M_{\rm ZAMS}$ as 

\begin{equation}
\frac{\omega_{\rm CHE}}{\omega _{\rm cr}} (Z, M) = e^{\alpha(Z)} M^{2}-e^{\beta(Z)} M+e^{\gamma(Z)} 
\label{eq: threshold_angular_velocity_single_star}
\end{equation}

\noindent where this form has been obtained by fitting the data in Fig. \ref{fig: Threhold_omega_Dutch_factor_variable}. The exponents are (see Appendix \ref{sec: Dutch_scaling_05} for the similar fit for a fixed value of Dutch scaling factor = 0.5)

\begin{equation}
\begin{array}{c}
\alpha(Z) = -1.692 \times 10^{7} Z^{3}+2.853 \times 10^{4} Z^{2}+571.5 Z -11.7 \\
\beta(Z)  = -3.391 \times 10^{7}  Z^{3}+1.602 \times 10^{5} Z^{2}+108.2 Z -6.01 \\
\gamma(Z) = -1.021 \times 10^{7} Z^{3}+5.753 \times 10^{4} Z^{2}+35.43 Z -0.565
\end{array}
\end{equation}

\begin{table}
	\centering
    \caption{Approximate mass range as a function of metallicity for the minimum threshold $\omega_{\rm CHE}/\omega_{\rm cr}$ shown in Eq. \ref{eq: threshold_angular_velocity_single_star} and Fig. \ref{fig: Threhold_omega_Dutch_factor_variable}. Note for $Z \in$ [10$^{-5}-0.002] \; M_{\rm min}$ near $\sim 15$ M$_{\odot}$ is not well reproduced by the fits. Hence we choose our $M_{\rm min}$ at the mid point of 15-20 M$_{\odot}$.}
    
    \label{tab: CHE_mass_range}
    \begin{tabular}{rrr}
    \hline
     $Z$    &     $M_{\rm min}$ [M$_{\odot}$] &  $M_{\rm max}$  [M$_{\odot}$]\\
    \hline \hline
    10$^{-5}-0.002$ & 15-20  & $M_{\rm ZAMS}$ as limit ${\omega_{\rm CHE}/\omega_{\rm cr} \rightarrow 1}$ \\
    $\gtrsim$ 0.003 & 30-35  & $M_{\rm ZAMS}$ as limit as ${\omega_{\rm CHE}/\omega_{\rm cr} \rightarrow 1}$ \\
    \hline
    \end{tabular}
\end{table}

The lower mass limit for CHE lies near $\sim 15$  M$_{\odot}$ ($\sim 30-35$  M$_{\odot}$ for $Z \geq 0.003$) and the upper limit is set as $\omega_{\rm CHE}/\omega_{\rm cr} \rightarrow 1$ (summarized in Table \ref{tab: CHE_mass_range}). The ratio $\omega_{\rm CHE} /\omega_{\rm cr}$ decreases with increasing mass and reaches a minimum before rising again. This is because as discussed earlier massive stars are easier to mix due to lower value of $\tau_{\rm mix}$ and $\tau_{\rm ES}$ compared to $\tau_{\rm MS}$.
On the other hand, as the mass increases so does the mass loss rate so there is a two-way competition between efficient mixing and specific AM loss (induced by mass loss) causing the parabolic shape.
Fig. \ref{fig: Threhold_omega_Dutch_factor_variable} also shows that stars with lower values of $Z$ have a larger $M_{\rm ZAMS}$ domain and are relatively easy to experience CHE. Due to growing AM loss in stellar winds, the $M_{\rm ZAMS}$ domain continuously shrinks with increasing $Z$ such that for $Z > [0.005-0.006]$ no CHE is possible (though this limit is a function of the wind mass loss rate, see Fig. \ref{fig: Threhold_omega_Dutch_factor_05}).



\section{The minimum threshold accretion for CHE} \label{sec: threshold_accretion_CHE}

\begin{figure*}
    \centering
    \vspace{-0.5cm}
    \includegraphics[width = 0.8\linewidth]{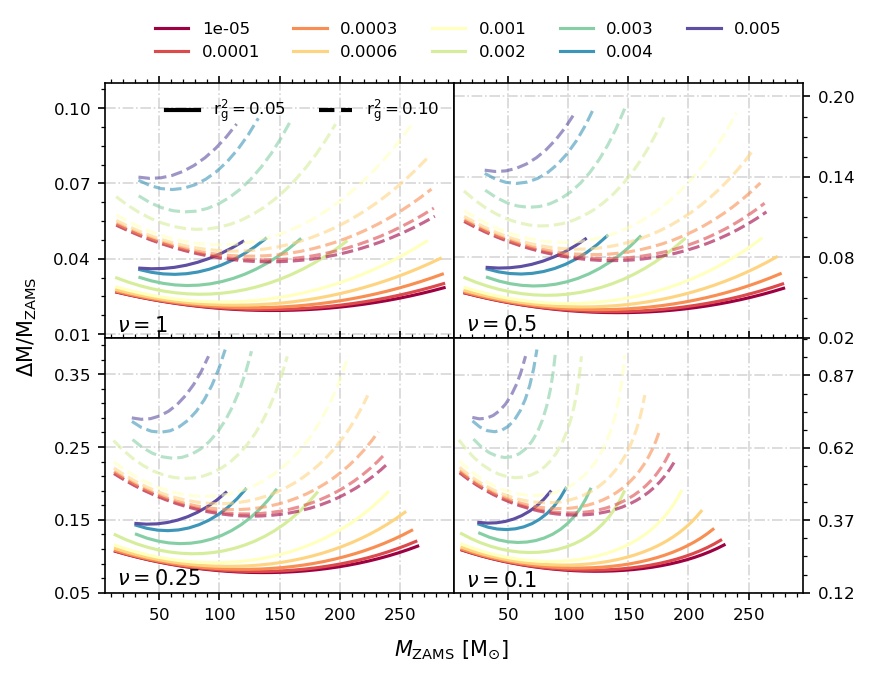}
    \vspace{-0.3cm}
    \caption{The minimum threshold accretion required by an accretor to experience CHE. Here r$_{\rm g}$ is the radius of gyration of the secondary at the onset of accretion. For demonstration we have chosen fixed values of $r_{\rm g}^{2} = 0.05, 0.1$. A lower value of $r_{\rm g}$ implies that star is more compact and vice versa. The curves have been derived from Eq. \ref{eq: Packet_modified}. The legend represents the initial metallicity of the star. Variable $\nu$ parameterizes the efficiency of accretion of AM from the Keplerian disk.}
    \label{fig: Acc_CHE}
\end{figure*}

Now that we know the $\omega_{\rm CHE}$ for a single star, can one recreate such rapidly rotating stars in a binary system? As hinted in section \ref{sec: intro} two such mechanisms have been proposed. Below we focus on accretion induced spin-up while the case of tidal CHE will be discussed in section \ref{sec: tidal CHE implementation}.

\subsection{Accretion induced spin-up}

During mass transfer between the binary components (of a close binary system) the infalling matter can gain a large amount of AM from the orbit \citep{Wilson1975}.
This matter eventually forms a Keplerian (critically rotating) disk around the secondary star from where it is then accreted.
In some cases, the accretion could instead be ballistic especially if the two stars are very close \citep{Lubow_1975} but for close binaries (although not close enough to experience tidal CHE) tides would be strong enough \citep{Hut_1981} to spin down the AM gaining accretor and hence a sustainable rapid rotation would not be realized.
As such we assume that all potential accretion CHE candidates are spun up due to accretion occurring only from the disk.
For systems forming accretion disks, \cite{Packet1981} showed that only a few per cent (5-10) of the accreted matter can cause the secondary to spin up to its critical velocity.
For this they assumed that the star is rotating as a solid body with a fixed radius and radius of gyration and the accretion happens from a disk near the equatorial plane with negligible viscosity. Their results showed that mass accreting stars can be easily spun up and therefore rotation cannot be ignored for accretors.

\textsc{bpass} has traditionally used the 5 per cent accretion threshold from \cite{Packet1981}, mass post accretion $ \geq 20$ M$_{\odot}$ and a metallicity value of $Z \leq 0.004$ ( \cite{Yoon_Langer2006_Xrbs}) as an indicator of the secondary experiencing CHE, although the relation in \cite{Packet1981} was derived for spinning up the star to its critical velocity.
In this way, one could underestimate the effect of accretion induced CHE as we are demanding more accretion than is formally required.
On the other hand, in certain places, one overestimates the effect due to not taking into account the metallicity dependence for a star on its ability to experience CHE as a function of its $M_{\rm ZAMS}$.
In particular, more massive stars with higher metallicity have stronger winds and therefore cannot experience CHE even if they are set to rotate rapidly at ZAMS. Therefore below we find the amount of minimum mass that needs to be accreted by the secondary to just become fully mixed both as a function of initial mass and metallicity. 

\subsection{Minimum accretion for CHE}

The below equation was derived by \cite{Packet1981} and is a measure of the fraction of accreted mass $\Delta M/M$ that would be sufficient to spin up the accretor star to its critical velocity (assuming solid body rotation) on the MS 

\begin{equation} 
\frac{\Delta M}{M} = \frac{1-\frac{3}{2} r_{g}^{2} - {\left( \left(\frac{3}{2} r_{g}^{2}-1\right)^{2}  -\left(2 r_{g}^{2}-1 \right) r_{g}^{2}\left(1- \frac{\omega_{\rm in}}{\omega_{\rm cr}}\right) \right)}^{1/2}} {r_{g}^{2} -\frac{1}{2}}
\label{eq: packet}
\end{equation} 

\noindent Here $r_{\rm g}, \: \omega_{\rm in}, \: \omega_{\rm cr}$ are the radius of gyration and the initial and critical angular velocity of the star respectively.
Though this expression is a function of $r_{\rm g}$ it does not directly depend on the mass and metallicity of the star.
In Appendix \ref{Sec1: Appendix} we derive a generalized version of this equation, which gives the minimum threshold mass accretion for a star to spin up just enough to experience CHE.
We assume that the accretion happens from the inner edge of the Keplerian disk and  that the matter being accreted is rotating in the same direction as the initial spin of the star.
Thus the equation takes the form


\begin{equation} \label{eq: Packet_modified}
\frac{\Delta M}{M} =  \sqrt{\frac{1}{4}\left(r_{\rm g}^{2} \frac{\omega_{\rm CHE}}{\nu \omega_{\rm cr}}+1\right)^{2}-r_{g}^{2} \frac{\omega_{\rm in}}{\nu \omega_{\rm cr}}}-\left(\frac{1}{2}-\frac{r_{\rm g}^{2} \omega_{\rm CHE}}{2 \nu \omega_{\rm cr}}\right)
\end{equation}

\noindent where $\nu$ is now a free parameter that controls the efficiency of accretion of AM  by the star from the Keplerian disk and $\omega_{\rm CHE}$ is the minimum angular velocity required by a star to experience CHE. Substituting $\omega_{\rm CHE}/\omega_{\rm cr}$ from Eq. \ref{eq: threshold_angular_velocity_single_star} in Eq. \ref{eq: Packet_modified} leads to a semi-analytical expression for the minimum threshold mass accretion required by a star to spin up just enough to experience CHE. Since Eq. \ref{eq: threshold_angular_velocity_single_star} is a function of $M_{\rm ZAMS}$ and $Z$ consequently is the expression in Eq. \ref{eq: Packet_modified}.



The $\omega_{\rm CHE}$ in Eq. \ref{eq: Packet_modified} can be replaced with any desired below critical value of $\omega_{\rm final}$ which will then result in the corresponding amount for $\Delta M / M$ required to be accreted to spin up the star to $\omega_{\rm final}$.  Substituting $\omega_{\rm CHE} = \omega_{\rm cr}$ and $\nu = 1$ in the Eq. \ref{eq: Packet_modified} reduces it to Eq. \ref{eq: packet} {\footnote{\cite{Packet1981} made further approximations and therefore the reduced equation is not in complete agreement with Eq. \ref{eq: packet}}}. Moreover if the angular velocity $\omega_{\rm in} = 0$ at the beginning of accretion, then Eq. \ref{eq: Packet_modified} reduces to 

\begin{equation} 
\frac{\Delta M}{M} = r_{g}^{2} \frac{\omega_{\rm CHE}}{\nu \omega_{ \rm cr}}
\label{eq: Packet_modified-with-omega=0}
\end{equation}

\noindent though this case is not particularly interesting as in most cases the accretor is assumed to have some $\omega_{\rm in}$ either due to AM acquired from the parental molecular cloud or due to tidal effects from a companion.

Fig. \ref{fig: Acc_CHE} plots Eq. \ref{eq:  Packet_modified} as a function of $M_{\rm ZAMS}$ and $Z$ with a conservative choice of $\omega_{\rm in} = 0$. For the sake of demonstration in Fig. \ref{fig: Acc_CHE} we pick $r_{\rm g}^{2} = 0.05, 0.1$ where a lower value of $r_{\rm g}$ implies that the star is more compact and vice versa. 
Later (section \ref{sec: Results}), the value of $r_{\rm g}$ will be derived from the structure of the \textsc{bpass} models.
We find that in contrast to the traditionally-assumed 5 per cent threshold accretion for spinning up the accretor star with $Z \lesssim 0.004 {\rm \; and \;}$ mass $\gtrsim 20$ M$_{\odot}$, this value can drop as low as 2 per cent for $r_{\rm g}^{2} \sim 0.04$.
Moreover, this is also now a function of $M_{\rm ZAMS}$ and $Z$, and therefore at certain places, one might even overestimate the efficiency of accretion induced spin up (e.g., see Fig. \ref{fig: Old_vs_new_accrt_CHE} later).

For stars with $Z < 0.001$, the minimum threshold accretion for CHE has a weaker dependence on metallicity for an extended range of $M_{\rm ZAMS}$, though the $M_{\rm ZAMS}$ range is dependent on the efficiency of accretion of AM from the Keplerian disk. On the other hand, for $Z > 0.005-0.006$ no amount of accretion can homogenize the star for an extended period of time.



\subsection{Accretion CHE versus CHE in isolated star} \label{sec: Accretion CHE versus CHE in isolated star}

One might ask, how valid is it to use Eq. \ref{eq: threshold_angular_velocity_single_star} (derived for single star models rapidly rotating at birth) to justify the case for accretion induced CHE?
After studying accretion induced CHE in a close binary system, \cite{Cantiello2007} concluded that the internal evolution of the mass-gaining star after accretion is almost identical to a rapidly rotating (at birth) single star.
In particular, both stars have similar tracks on the HR diagram and similar internal profiles on the Kippenhahn diagram, and both experience CHE. They also speculated that the metallicity threshold for CHE via accretion should be similar as that for CHE for single stars. 

The downside is that the study contains only one model at a specific metallicity. Nonetheless, these massive MS stars form a homologous sequence so extending their conclusion to other MS models would still be a fair approximation. They further assumed that the specific AM gained by the accretor is equivalent to the specific AM of the accretor's surface.
On the other hand, a star accreting from a Keplerian disk would in principle need to accrete relatively less mass to spin up. Since the efficiency of accretion of AM remains uncertain, here the free parameter $\nu$ (see, Eq. \ref{eq: Packet_modified}) controls the efficiency at which the AM is accreted by the star from the Keplerian disk. A more detailed numerical analysis of this aspect is outside the scope of current work and might be of future interest.

\section{CHE implementation in BPASS} \label{sec: CHE in BPASS}

\subsection{BPASS input parameters}

To conduct a population study, we use \textsc{bpass} v2.2.1 stellar models \citep{Bpass2017, Bpass_2018} which provide detailed pre-processed single and binary star models computed up till the time of core carbon exhaustion followed by core-collapse SN (for the mass range applicable here). Motivated by \cite{Moe_Di_Stefano2017}, the initial binary period ($P$) distribution  is taken as log($P$/days) $\in$ [0,4] separated by increments of 0.2 dex and initial binary mass ratio $q_{i}  \in [0.1, 0.9]$ with 0.1 step increment. The orbit of the binaries is assumed to be circular at birth.

\begin{figure*}
    \centering
    \vspace{-0.25cm}
    \includegraphics[width = 0.65\linewidth]{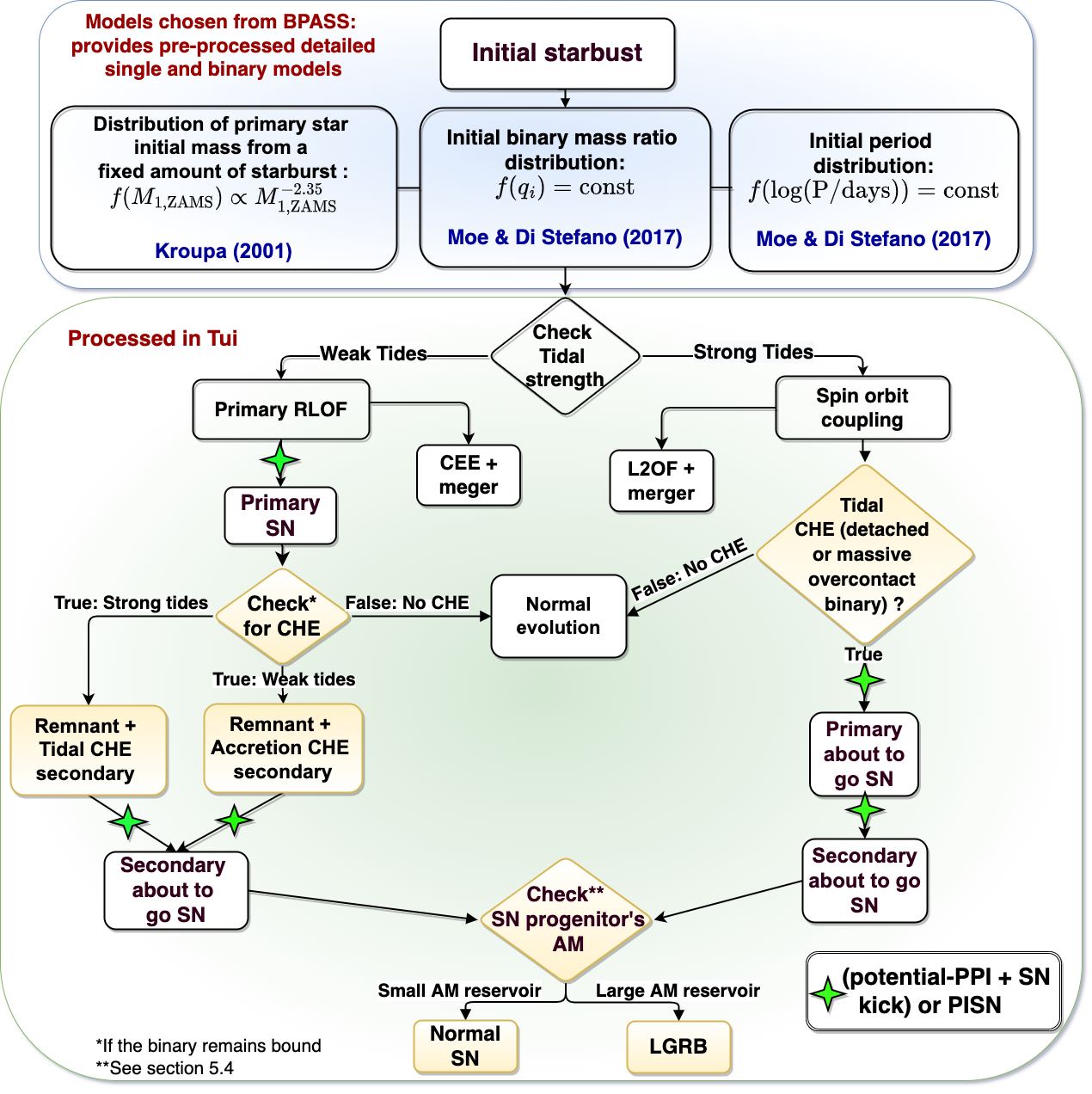}
    \caption{A schematic diagram showing the implementation conducted in section \ref{sec: CHE in BPASS}.  The yellow boxes are where additions/modifications have been made in the current work. The green star represents the potential occurrence of PPI, SN kick or PISN. Although the green  star is drawn before SN, the kick should occur post SN. The diagram only focuses on the core processes in single and binary stellar evolution that are relevant to the current work.}
    \label{fig: Schematic}
\end{figure*}

We use the \cite{Kroupa2001} initial mass function to determine the number of primaries that form at a given initial mass and focus only on massive stars with the allowed primary initial mass ranging from  [10-300] M$_{\odot}$.
The allowed initial metallicities are $Z = 10^{-5}$, $10^{-4}$, 0.001, 0.002, 0.003, 0.004, 0.005, 0.006, 0.008, 0.01. \textsc{bpass} uses the Dutch stellar wind mass-loss prescription as a combination of \cite{deJager_1988, Vink_2001} and \cite{Nugis_Lamers_2000} depending on the evolutionary stage of the star. We refer the reader to Fig. \ref{fig: Schematic} and the instrument paper (\citealt{Bpass2017}, section 2.2 therein) for more information.


In this work we use a satellite routine of the \textsc{bpass} suite (\textsc{tui} v1) to make additions in a post-processing manner (discussed below and a schematic is presented in Fig. \ref{fig: Schematic}) to the \textsc{bpass} models. \textsc{tui} was previously used in \cite{Ghodla2022}, but we name it here for the first time. Apart from the implementation discussed below (section \ref{sec: tidal CHE implementation}, \ref{sec: implementaion of accretion CHE}), we also account for the occurrence of pulsation pair-instability (PPI - see \citealt{Woosley_PPISNe_2007}) within the \textsc{bpass} models in \textsc{tui} by following the fits in \cite{Stevenson_2019} - Eq. 4 therein - for the non rotating helium star models of \cite{Marchant_2019_PPISne}.

\subsection{Implementation of tidal CHE} \label{sec: tidal CHE implementation}

Tidal friction in tight binaries is very effective in aligning, circularising and synchronizing the orbital period of the stars with their spin period (\citealt{Hut_1981}).
Hence, in \textsc{tui}, the potential tidal CHE candidates are assumed to be tidally locked at birth making their spin angular velocity equal to their orbital velocity.
These stars should undergo more efficient mixing than single stars with the same initial angular velocity because orbital AM is continuously fed into them as they spin down due to mass loss.
Hence we do not use Eq. \ref{eq: threshold_angular_velocity_single_star} but instead compare the resulting angular velocity to the threshold angular velocity for CHE as developed in \cite{Riley_2021} which was specifically developed for tidally locked stars (Appendix A therein). Their relation was provided for $10^{-4} \leq Z \leq 0.01$ but we extrapolate it to $Z = 10^{-5}$. If CHE is allowed, then we evolve the star homogeneously.

We continuously check for the fulfillment of CHE conditions while the surface helium mass fraction of the star is lower than 0.8. In the meantime, if the orbit widens due to mass loss, then that would quickly lead to reduction of the angular velocity of the star such that it could no longer experience CHE.
In most cases, within \textsc{tui} the latter would lead to a merger of the tight binary system due to subsequent Roche lobe overflow (RLOF).
If the star fails to fulfill the CHE condition and the binary does not merge, then we evolve the star as a normal star from its ZAMS. This is due to \textsc{bpass} having detailed pre-processed models which makes it inefficient and computationally intensive to first calculate CHE and then the normal phase of the evolution conjointly. 

\cite{Marchant2016} found that the stars that experience RLOF at ZAMS could still maintain co-rotation as long as their mass does not overflow the L2 point. This leads the stars to swap material and hence equalize their mass. To accommodate the findings of \cite{Marchant2016}, we equalize the mass of all systems that experience RLOF at ZAMS.
Here we follow \cite{Riley_2021} and assume that if the sum of the unperturbed radii of the stars $R_{1, \rm ZAMS} + R_{2, \rm ZAMS} < a$, then the system can avoid prompt merger, where $a$ is the initial separation between the binaries. Assuming conservation of orbital AM (with the stars carrying negligible AM) gives the new separation of the binaries as
\begin{equation}
	a^{\prime}=\left(\frac{4 M_{1} M_{2}}{\left(M_{1}+M_{2}\right)^{2}}\right)^{2} a
\end{equation}

To account for orbital widening due to mass loss in a system with at least one CHE component we use the Jeans mode of mass loss, i.e. assume that the winds are fast and isotropic and hence do not interact with the orbit and take away only the specific orbital AM of the mass losing star.  For circular orbits and  $|dM| \ll  M_{1,2}$  this can be approximated as (see Eq. 16 in \citealt{Hurley2002}, also see \citealt{deMink_Mandel2016}), $da = a |d M| / (M_{1}+M_{2})$ where $dM$ is the mass lost. \cite{deMink_Mandel2016} argue that in the case of tight binaries, for wind speeds comparable to the orbit speed of the stars, winds might even harden the orbit or diminish orbital widening.
Here we do not take this into account. We use the Hobbs kick prescription (\citealt{Hobbs_2005}) for determining the SNe kick velocity and allow for another episode of tidal locking if the kick subsequently results in the formation of a tight binary but now with a compact remnant and a normal secondary star.

For determining the mass of the compact remnant, we use the standard remnant mass implementation in \textsc{bpass}, (see, \citealt{Ghodla2022}).
Finally, \textsc{bpass} does not yet evolve systems with $q_{i} = 1$ apart from the new post-processed addition which forces the stars in massive over-contact binaries to equalize their mass at birth. So we could potentially be missing some equal mass components that would have experienced CHE without RLOF at ZAMS.

\subsection{Implementation of accretion CHE} \label{sec: implementaion of accretion CHE}

For implementing accretion induced CHE, we use Eq. \ref{eq: Packet_modified} and set $\omega_{\rm in} = 0$, conservatively assuming that the accretors in \textsc{bpass} are not initially rotating, which then reduces it to Eq. \ref{eq: Packet_modified-with-omega=0}.
At the beginning of the mass transfer, our models have a range of values for $r_{\rm g}^{2}$, spanning [0.025-0.085] with the mean lying around 0.049 and a standard deviation of 0.013. We note that $r_{\rm g}^{2}$ decreases as the star evolves along the MS due to the matter becoming more concentrated near the center.
Hence we do not consider stars with $r_{\rm g}^{2} \leq 0.025$ as potential CHE candidates as these systems would have evolved too far along their MS. The convective core of such systems would not have enough time to adapt to the increased mass.
As a result rejuvenation of the star due to mixing would be altogether avoided (e.g. \citealt{Braun_Langer1995}).


In sufficiently close binaries,  the differential gravitational pull of one star on the other can excite oscillations causing dissipation and redistribution of energy and AM due to tidal deformations. 
Here we allow for tidal spin-down of the mass accreted secondary (disfavoring its prospects for experiencing accretion CHE) due to dissipation processes like turbulent viscosity with a synchronization time scale $\tau_{\rm sync}$ as in \cite{Zahn_1977_Tidal_sync}:

\begin{equation}
    \tau_{\rm sync} \approx q^{-2}(a/R)^{6} \; {\rm year}
    \label{eq: t_turb}
\end{equation}

\noindent where $q$ is the mass ratio $M_{2}/M_{1}$ and star with mass $M_{2}$ is the one for which Eq. \ref{eq: t_turb} is valid, $R$ is the radius of $M_{2}$ and $a$ the separation between the stars. The eccentricity of the binary is considered negligible.

We only initiate tidal synchronization once the mass transfer phase is over and hence $M_{2}$ and $M_{1}$ are chosen accordingly.
Also, the time frame for tidal spin down is limited by the MS lifetime of the accretor. The above $\tau_{\rm sync}$ was derived for stars with convective envelopes while stars in the mass range considered here should have radiative envelopes.
Nevertheless, for fast rotators, \cite{Tolendano2007} found that even in stars with radiative envelopes,  differential rotation induces turbulence in the surface layers of the star such that the $\tau_{\rm sync}$ timescale in Eq. \ref{eq: t_turb} remains valid.



\section{Observational implications} \label{sec: Results}

To study the observational implications of CHE, we evolve the \textsc{mesa} models till core carbon depletion. Due to numerical reasons, the lowest metallicity models halt when core carbon fraction $\leq 0.2$ (see Fig. \ref{fig: surf_he_cntr_c12}). The most significant change in $J_{\rm total}$ occurs prior to core helium depletion after which $J_{\rm total}$ only experiences a minor variation. 
Post core carbon depletion, we expect the change in net AM of the CH stars to be minimal. Hence we approximate the pre core collapse total AM content of the star with its AM at carbon depletion. Our CH \textsc{mesa} models do not account for mass and AM loss induced due to PPI since our models are not evolved far enough to experience PPI. 
Nevertheless, PPI would likely not affect the specific AM carried by the inner region of these compact stars which is crucial for the following analysis. Lastly, we follow \cite{Woosley_2017} and assume that all stars with final helium core mass $M_{\rm He}$ in the range $65{\rm M}_{\odot} \lesssim M_{\rm He} \lesssim 135{\rm M}_{\odot}$ undergo pair-instability SN (PISN - \citealt{Fowler_and_Hoyle1964}) hence leaving no remnant. From here on, we do not study CH models above the PISN mass range.

Below we show some observational implications of the CH stars considered here including their frequency of occurrence and their contribution to ionizing photons production and energetic transients like SLSN-I, GRBs, Ic-BL SN.
For this (excluding section \ref{sec: ionizing_photons}), we only consider accretion-induced CH systems because it is unclear as to how much AM would be retained in a tidal CH star by core-collapse.
This is because unlike a CH star spun up by accretion, which essentially acts as an isolated system post mass transfer (unless the binaries are close enough leading to tidal spin down, see Fig. \ref{fig: CHE_percent}), the AM content of a tidal CH star is dictated by the influence of tides which get weaker as the binary's orbit expands due to mass loss.
Moreover from Fig. \ref{fig: CHE_percent}, we find that depending on the metallicity, the frequency of stars experiencing accretion CHE dominate their tidal CHE counterpart by over an order of magnitude making them less important for the following study.




\subsection{The frequency of chemically homogeneous stars}

\begin{figure}
    \centering
    \vspace{-0.5cm}
    \includegraphics[width = 1.02\linewidth]{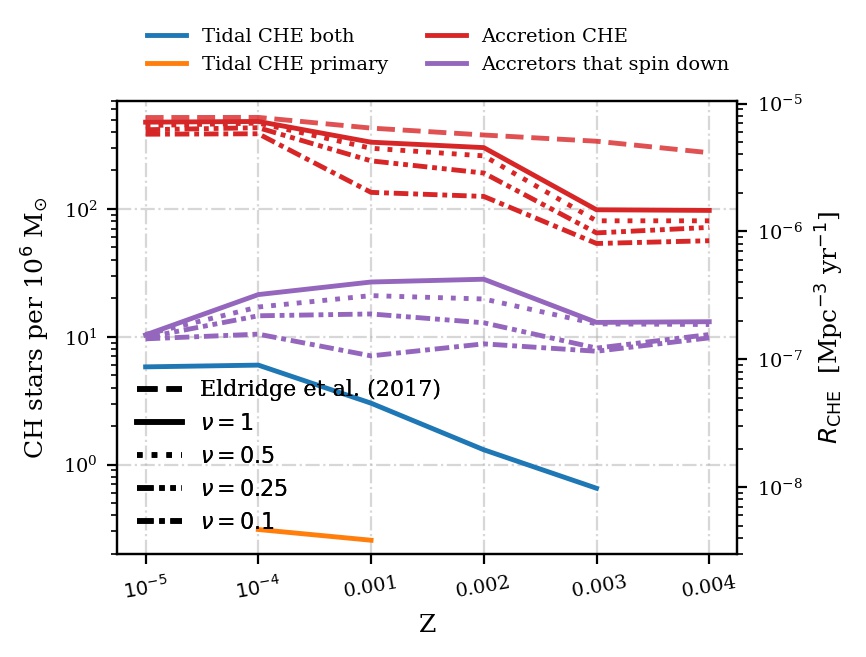}
    \vspace{-0.5cm}
    \caption{The number of chemically homogeneous systems as a function of $Z$ from an initial starburst of 10$^{6}$ M$_{\odot}$.
    The results from the previously used accretion CHE assumption in \textsc{bpass} has been shown with the label \protect\cite{Bpass2017}.
    The orange and blue line shows the number of tidal CH systems where either both or only the primary star undergoes CHE respectively.
    The purple lines represent those systems that accreted enough AM to experience accretion CHE but were close enough to the companion to undergo tidal induced spin down. Variable $\nu$ is a free parameter that models the efficiency of AM accretion from the Keplerian disk onto the accretor.
    On the RHS of the y-axis are their local rates of formation in Mpc$^{-3}$ yr$^{-1}$.
    Star formation at $Z < 0.001$ is unlikely to dominate in the local Universe (i.e. at redshift zero) so the rate mainly at larger values of $Z$ should be looked at.
    Additionally, contrary to our assumption, not all the star formation would occur at a single $Z$. Therefore the value $R_{\rm CHE}$ should be lowered roughly by an order of magnitude if we are concerned with the rate at only one specific $Z$.}
    \label{fig: CHE_percent}
\end{figure}

\begin{figure*}
    \centering
    \vspace{-0.4cm}
    \includegraphics[width = 0.75\linewidth]{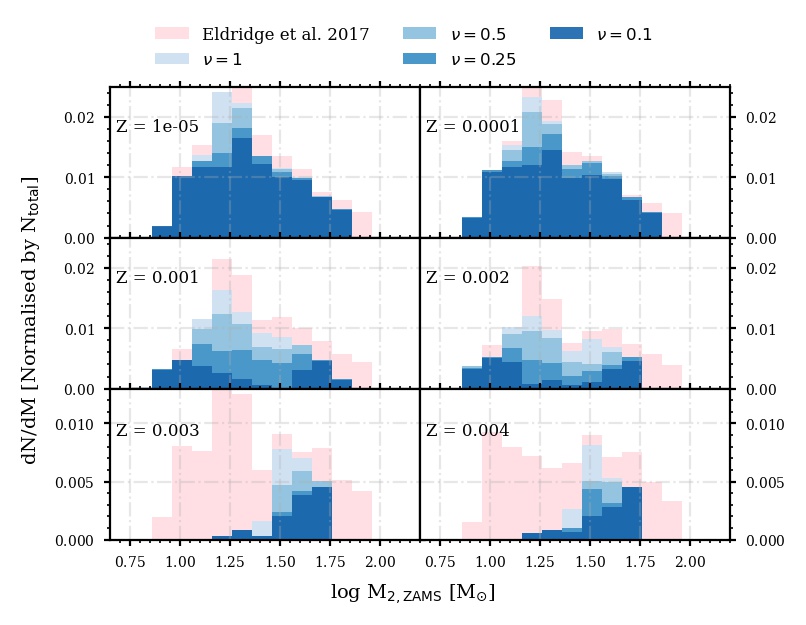}
    \vspace{-0.3cm}
    \caption{The variation in the mass distribution of the candidates that experience accretion CHE due to the old and the new formulation within the \textsc{bpass} population. With increasing $Z$, a depreciation in $\nu$ causes a noticeable change in the frequency of CH systems with the lower mass accretors now being disfavoured. We note that Eq. \ref{eq: threshold_angular_velocity_single_star} applies for the mass of the accretor at the end of the accretion phase and not $M_{\rm 2, ZAMS}$.}
    \label{fig: Old_vs_new_accrt_CHE}
\end{figure*}

We find that the upper bound on metallicity for CHE is a function of the physical process causing it. In particular, depending on the WR mass loss rate, Eq. \ref{eq: threshold_angular_velocity_single_star} (which is valid for an isolated rotating star and therefore for accretion CHE, see section \ref{sec: Accretion CHE versus CHE in isolated star}) has an upper limit near $Z \sim 0.005-0.006$. For the case of tidal CHE, the net AM content of the star is primarily dictated by spin-orbit coupling.
Therefore Eq. \ref{eq: threshold_angular_velocity_single_star} cannot directly be used in the same fashion here. Fig. \ref{fig: CHE_percent} displays the results of our population synthesis using \textsc{bpass} and shows that for tidal CHE the upper limit for homogeneous evolution lies near $Z \sim 0.003$.
As $Z$ increases, it becomes increasingly harder for both stars in the binary to experience tidal CHE till the end of MS. 

Most models initially undergo tidal CHE but soon drop out due to orbital widening induced tidal spin down.
Although initially, some secondaries do experience tidal CHE, none could successfully remain CH till the end of MS (this includes secondaries under the tidal influence of compact remnants).
These limits are strongly dependent on the rate of wind-induced AM loss and hence could change if a modification is made to the wind scheme. Additionally, we could be under-predicting the tidal CHE systems by a factor of few.
This is because in \textsc{bpass} the minimum initial period $P_{\rm ZAMS}$ is set to one day but other studies find numerous potential tidal CHE systems for $P_{\rm ZAMS} \lesssim 1$ day \citep{deMink_Mandel2016, Marchant2016, du_Buisson2020, Riley_2021}.
We note that as of now, the smallest  detected period for a potential CHE system belongs to VFTS 352 - a massive binary system in 30 Doradus region - with a period of 1.124 days \citep{Almeida_2015_Obs_CHE}.

Fig. \ref{fig: CHE_percent} shows the frequency of CH systems for our models as a function of $Z$. Here accretion CHE appears to be the dominating channel for producing homogeneous stars.  The frequency of CH systems decreases with increasing metallicity by a factor of a few.
Additionally, at most roughly one in ten potential accretion CH candidates experiences tidal spin down.
For the case of spin-orbit coupling, tidal CHE due to mass equalization in massive over-contact binaries dominates over single component tidal CHE.
If accretion quickly spins up the accretor's surface to critical rotation, then the further accreted mass would not contribute towards the AM of the accretor unless the surface AM is transferred inwards.

To calculate the efficiency at which the accreted matter can transport AM from the surface to the inner regions would require a full 3D magneto-hydrodynamical study. On the other hand, if the star’s surface layer on accretion quickly reaches breakup then the efficiency at which AM could be further accreted by the surface of the star from the inner edge of the disk would depend on the efficiency at which this accreted AM is transported inwards. 
For simplicity, we introduce an AM accretion efficiency parameter $\nu$ (see section \ref{Sec1: Appendix} in appendix) and show its influence on accretion CH systems in Fig. \ref{fig: CHE_percent}. Limiting the accretion efficiency of AM to the parameter $\nu$ would be equivalent to assuming that the rate of inward redistribution of AM  from the equatorial critically rotating surface layer is also equal to $\nu$.
We find that a low value of $\nu$ leads to a decrease in the number of accretion CHE systems by a factor of few and mostly at larger $Z$ values.
On comparing with the earlier \textsc{bpass} result for accretion CHE we find that the new relationship for CHE affects the frequency and distribution of CH systems in our stellar evolution code by a factor of few.
Fig. \ref{fig: Old_vs_new_accrt_CHE} shows the variation in accretion CHE rate due to the old and the new assumptions in the $M_{\rm ZAMS}$ space.
With increasing $Z$, a decrease in $\nu$ causes a noticeable change in the frequency of CH systems with the lower mass accretors now being disfavoured.


\subsection{Contribution towards ionizing photons} \label{sec: ionizing_photons}

Several studies (e.g. \citealt{Stanway2014, Stanway2016, Xiao_2018, Rosdahl_2018, Gotberg_2020}) have looked into the effects of binary interactions on the emitted spectra of the underlying stellar population. They conclude that binary interactions are crucial for the reionization of the Universe. But binaries alone cannot fully explain the observed strength of He II, C IV spectral lines of high redshift galaxies \citep{Eldridge_Stanway_2012, Stanway_2019}. This issue can be resolved if one includes the contribution of CH stars towards the ionizing photon density.

\subsubsection{Calculating emission rate of hydrogen ionizing photons}

Due to a high surface temperature and luminosity, the CH stars emit a relatively large fraction of their light in the far ultraviolet spectrum. These photons with wavelength $\lambda < 912$\AA \; have energies sufficient to ionize hydrogen atoms, producing luminous HII regions.
Although photons with smaller $\lambda$ could be absorbed by helium, these would eventually be re-emitted and hence one can integrate all the way from $\lambda \in (0, 912)$ \AA \; to calculate the net number of hydrogen ionizing photons.
Here we study the number of such hydrogen ionizing photons per second being emitted (by stars experiencing CHE) from a continuous star formation of 1 M$_{\odot}$ yr$^{-1}$ for a time period of 100 Myrs as follows

\begin{equation}
    N_{\rm \gamma} = \frac{1}{t_{\rm max}} \int_{0}^{t_{\rm max}} \Psi(t) \left( \int_{0}^{t_{\rm max} - t} Q_{\rm H}(t^{\prime}) dt^{\prime} \right ) dt
    \label{eq: Ionizing_photons_rate}
\end{equation}

\noindent where $t_{\rm max}$ is taken as 100 Myrs, $\Psi$ is the star formation rate and $Q_{\rm H}$ is the ionizing photons emission rate per unit of star formation as a function of time.
The ionizing photons production rate, $Q_{\rm H}$ is evaluated for the CH \textsc{mesa} models by calculating their stellar atmosphere structure employing the approach detailed in \cite{Bpass2017, Bpass_2018}. 

\begin{figure}
    \centering
    \includegraphics[width = 1\linewidth]{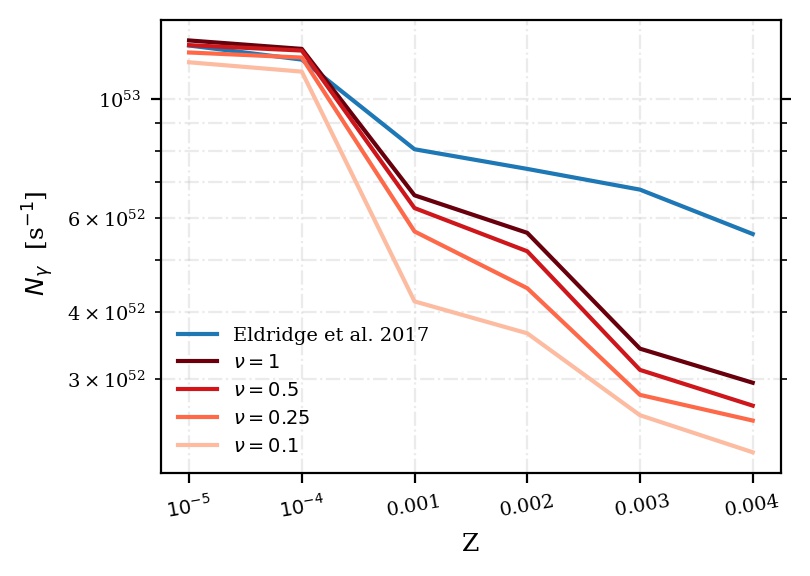}
    \vspace{-0.5cm}
    \caption{ The number of photons emitted ($N_{\gamma}$) with $\lambda < 912$\AA. Values are quoted assuming a constant star formation of 1 M$_{\odot}$yr$^{-1}$ for a period of 100 Myrs (see Eq. \ref{eq: Ionizing_photons_rate}). }
    \label{fig: ionising_photons_count}
\end{figure}

The $Q_{\rm H}$ resulting from these \textsc{mesa} models is in relatively good but not full agreement with the earlier CH \textsc{bpass} models in \cite{Bpass2017}. The CH \textsc{bpass} models considered in \cite{Bpass2017} were forced to remain fully mixed over the course of their lifetime while the CH \textsc{mesa} models continuously solve for the stellar structure to calculate the internal mixing evolution.
Consequently, the \textsc{mesa} models are more accurate and have longer lifetime when compared to their \textsc{bpass} counterpart. They also have a better time resolution. We note that unlike the period and mass ratio distribution used in \cite{Bpass2017} - forcing all the stars to be in binaries, potentially resulting in a larger population of mass transferring components - here we rely on the \cite{Moe_Di_Stefano2017} period and mass ratio distribution for all our calculation (this also holds for result referred with the label \cite{Bpass2017} in Fig. \ref{fig: ionising_photons_count}).




\subsubsection{Fewer ionizing photons at larger metallicities}

We find that for a given starburst while the emission of ionizing photons from the tidal CH stars dominates until $\sim 6$ Myrs, the accretion induced CH star could keep on emitting such photons until $\sim 32$ Myrs from the onset of star formation.
Fig. \ref{fig: ionising_photons_count} shows our result for the ionizing photon count. The emission rate approximately matches the earlier findings (\citealt{Bpass2017}) at low values of $Z$ (Note that \citealt{Bpass2017} did not account for tidal CHE).
For $Z \gtrsim 0.001$  deviation from \cite{Bpass2017} ($\sim$ 15-50 per cent) can be seen especially for low values of $\nu$. For $Z> $ 0.002-0.003 decreasing $\nu$ considerably lowers the emission rate anywhere between $\sim$ 40-65 per cent.
This is to be expected as the current study narrows the span of the $M_{\rm ZAMS}$ range over which a star at larger $Z$ can experience CHE more so for $Z \gtrsim 0.001$ (see Fig. \ref{fig: Old_vs_new_accrt_CHE}).
\cite{Rosdahl_2018} found that the \textsc{bpass} stellar population reionizes the Universe a little faster (near redshift $z \sim 7$, however also see section 3.3 in \citealt{Rosdahl_2022}) than  predicted by the model-dependent observations (near $z \sim 6$).
They attribute this disagreement to  the higher escape fractions and higher ionizing photon flux produced by the \textsc{bpass} population. The reduction in $N_{\gamma}$ as shown here could revise their estimates and hence help in rectifying part of this disagreement.

\subsection{Contribution towards SLSNe-I and Ic-BL with a magnetar as the source} \label{subsec: Magnetar_implication}

Energetic Supernovae such as SN 2005ap \citep{Quimby_2007}, SCP 06F6 \citep{Barbary_2009}, SN 2009jh \citep{Drake_2009}, SN 2010gx (\citealt{Pastorello_2010_1}, for more events see  \citealt{Nicholl_2017, Villar_2018, Blanchard_2020}), are reported to be more than ten times brighter than most type Ia SN and lack traces of hydrogen in their spectra (see \citealt{Moriya_2018, Inserra_2019} for a review).
Due to the presence of these and a few other characteristics they have been classified into a new class of SN called superluminous SN type I (SLSN-I) (\citealt{Quimby_2011, Gal_Yam_2012}). They are believed to be caused either due to PPI (e.g. \citealt{Gal_Yam_2009}) leading to the reduction of pressure due to $e^{-}e^{+}$ pair production in the cores of massive evolved stars with pre core collapse helium core mass in  $40{\rm M}_{\odot} \lesssim M_{\rm He} \lesssim 60{\rm M}_{\odot}$ \citep{Woosley_PPISNe_2007} - though, \citealt{Woosley_2017} found that secondary energy sources might be required - or due to additional envelope energy  injection by a central engine following core-collapse SN.
This energy could be sourced from the creation and spin-down of a newly-formed magnetar \citep{Woosley_2010, Kasen_Bildsten_2010} or a collapsar-like accreting BH leading to the formation of relativistic jets \citep{Woosley1993_collapsar, Woosley_1999}.

\begin{figure}
    \centering
     \vspace{-0.5cm}
    \includegraphics[width = 1\linewidth]{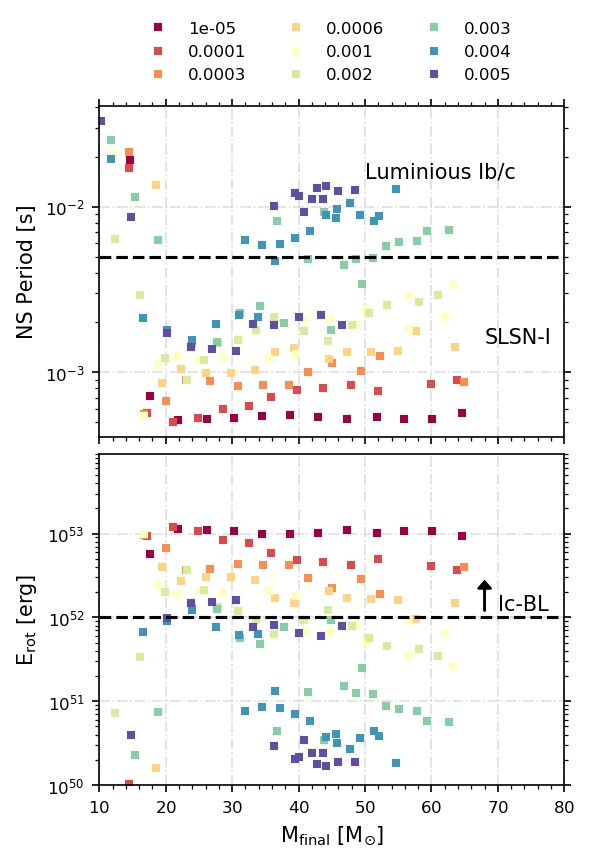}
    \vspace{-0.5cm}
    \caption{Figure assumes all progenitors collapse to produce a NS. In the top sub-figure, the lines divide the luminous Ib/c SNe from the SLSNe-I assuming that a magnetar of period $P_{\rm NS} \lesssim 5$ ms would be required to produce a SLSN-I. In the bottom sub-figure the systems falling above the $E_{\rm rot} > 10^{52}$ erg are considered as possible candidates for Ic-BL (see, section \ref{sec: Estimating peiod and E_rot} for details). Since most of the highest mass stars would likely form a BH, we expect fewer systems producing SLSN-I. On the other hand, Ic-BL can be produced even if a BH is formed (via collapsar mechanism, Fig. \ref{fig: Eiso_Ewind_vs_Mfinal}). Some stars with $Z = 0.004, 0.005$ fail during core helium burning (see Fig. \ref{fig: surf_he_cntr_c12}). On complete evolution, these systems should not hold more AM than those at $Z = 0.003$. Consequently, such stars should not feature in the SLSN-I/Ic-BL region.}
    \label{fig: P_NS_E_rot_vs_Mfinal}
\end{figure}

On the other hand the broadlined SN type Ic (hypernova, Ic-BL) possesses unusually broad spectral lines indicating  extremely high expansion velocity (15,000--30,000 km s$^{-1}$, e.g. \citealt{Modjaz_2016}). Some recent work has shown a connection between SLSN-I and Ic-BL (e.g. \citealt{Pastorello_2010, Metzger_2011, Metzger_2015, Blanchard_2019, Shankar_2021, Gomez2022}), speculating that a common central engine might be behind both.
Though at present this remains inconclusive, here (without making an assumption as to how a central engine could reconcile SLSNe and Ic-BL into a single framework) we show that purely on the energetic grounds, a magnetar born from a CH star offers scope to simultaneously act as a progenitor of both.

Fig. \ref{fig: P_NS_E_rot_vs_Mfinal} and \ref{fig: P_NS_Erot_MZAMS} shows the period $P_{\rm NS}$ and rotational energy $E_{\rm rot}$ of the neutron star formed from the single star CH models considered here. Here we  choose $P_{\rm NS} = 5$ ms as the lower bound on $P_{\rm NS}$ for a successful SLSN-I though $P_{\rm NS}$ as large as $7.68_{-0.9}^{1.37}$ ms have been inferred (event PTF11hrq, \citealt{Villar_2018}). For successful Ic-BL we assume that $E_{\rm rot} > 10^{52}$ erg (e.g. \citealt{Metzger_2011, Shankar_2021}). Under favorable conditions (see below) many of the CH stars considered here have suitable traits to act as the progenitors of SLSN-I and Ic-BL.

\subsubsection{Estimating spin period and rotational energy of the Neutron star} \label{sec: Estimating peiod and E_rot}

Similar to \cite{Fuller_2022}, we take the baryonic mass of the NS as $1.6 {\rm \; M}_{\odot}$ (which results in a gravitational masses close to 1.4 M$_{\odot}$) and its moment of inertia as $I_{\rm NS} = 1.5 \times 10^{45}$ g cm$^{2}$ (a suitable value for a rapidly rotating NS, see \citealt{Worley_2008}).
We also assume that the AM ($J_{\rm core}$) contained within the inner 1.6 ${\rm \; M}_{\odot}$ is conserved during the phase of core-collapse which then gives the period of the NS as $P_{\mathrm{NS}}=2 \pi I_{\mathrm{NS}} / J_{\mathrm{core}}$ and the corresponding rotational energy as $E_{\rm rot} = J_{\rm core}^{2} / 2I_{\rm NS}$.
Increasing the baryonic mass to an extreme value of 3 M$_{\odot}$ (and $I_{\rm NS} \sim 3.5 \times 10^{45}$ g cm$^{2}$ - \citealt{Worley_2008}) would also increase the value of $J_{\rm core}$, consequently this would roughly half the value of $P_{\rm NS}$ and increases $E_{\rm rot}$ by a factor of $\sim 4$ in Fig \ref{fig: P_NS_E_rot_vs_Mfinal}.

\subsubsection{Uncertainties and additional assumptions}

Realistically, a fraction of AM contained in the  core ($J_{\rm core}$) would also be carried away by neutrinos during the formation of a proto NS. 
\cite{Baumgarte_Shapiro_1998} showed that neutrinos are not very efficient in removing AM and \cite{Janka_2004} concluded that at most they can remove 43 per cent of the total AM of the NS. Here we conservatively assume full efficiency of AM removal by neutrinos.

In Fig. (\ref{fig: P_NS_E_rot_vs_Mfinal}, \ref{fig: P_NS_Erot_MZAMS}) we force all systems to form NS.
This is certainly not true but uncertainty regarding the maximum pre-explosion mass of NS forming progenitor still remains.
In particular, so called islands of explodability have been found in numerical models of non-rotating systems implying that a simple progenitor mass division for NS or BH formation might not exist \citep{Ertl_2016, Sukhbold_2016, Ertl_2020}. On calculating the ejecta mass for 62 different SLSN events using light curve models, \cite{Blanchard_2020} inferred that SLSN (with a magnetar as the source) could occur even for pre-explosion mass $\sim$ 40 M$_{\odot}$.
More support comes from \cite{Aguilera-Dena_2020} who studied the explodeability of CH stars and found that a NS could form for pre-collapse mass $\lesssim 30$ M$_{\odot}$.
In the current work, for $M_{\rm ZAMS} > 45$ M$_{\odot}$, we expect PPI to shed a fraction of the star's mass prior to core collapse.
Hence $M_{\rm final}$ would not be as large as shown in Fig. \ref{fig: P_NS_E_rot_vs_Mfinal} once PPI is considered. We suppose that if PPI occurs, it would not much affect the core angular momentum of the CH star. Additionally for rotating stars if $E_{\rm rot}$ > $E_{\rm bind}$ (where $E_{\rm bind}$ is the progenitor's envelope binding energy), depending on the efficiency of coupling of $E_{\rm rot}$ to the SN shock, a NS could in principle form, even from very massive stars (e.g. \citealt{Metzger_2011}).

The minimum value of $P_{\rm NS}$ in Fig. \ref{fig: P_NS_E_rot_vs_Mfinal} is $\sim 0.4$ ms.
This is smaller than the fastest-spinning NS detected  with a frequency of 716 Hz ($P_{\rm NS} \sim 1.4$ ms, \citealt{Hessels_2006}).
Studies place a lower bound on the $P_{\rm NS}$ such that $P_{\rm NS} \gtrsim 1$ ms once the NS settles into a lepton poor state \citep{Strobel_1999} and further decrement in $P_{\rm NS}$ could only occur via subsequent accretion. Us underestimating the value of $P_{\rm NS}$ arises due to not following the core collapse and the subsequent NS formation in detail.
Studies (e.g. \citealt{Ott_2006} and reference therein) propose mechanisms such as non-axisymmetric rotational hydrodynamic instabilities that could spin down rapidly rotating NSs. Nevertheless, even with a slight increment in $P_{\rm NS}$, these systems would most likely still remain in the SLSN-I/Ic-BL domain.
Although these instabilities might also change the period of NSs with a larger value of $P_{\rm NS}$ consequently removing some of them from the SLSN-I/Ic-BL domain. 

For Ic-BL the lower limit of $E_{\rm rot} \sim 10^{52}$ erg might not be enough to accelerate the entirety of the ejecta to velocities above 15,000 km s$^{-1}$.
Assuming the ejecta to have kinetic energy $E_{\rm ej} = 0.5 M_{\rm ej}v_{\rm ej}^{2}$, then at most an ejecta mass of 5 M$_{\odot}$ can be accelerated to $v_{\rm ej} \sim$ 15,000 km s$^{-1}$ (under full efficiency of conversion of $E_{\rm rot}$ into $E_{\rm ej}$, cf. \citealt{Fuller_2022}).
Along similar lines, $M_{\rm ej} = 20$ M$_{\odot}$ would require $E_{\rm rot} = 4.5 \times 10^{52}$ erg meaning only the lowest $Z$ stars would be able to experience Ic-BL.
Alternatively, it could be that only a fraction of $M_{\rm ej}$ needs to be accelerated to large velocities to make the source appear broadlined, in which case the high $Z$ CH stars can also contribute towards Ic-BL for large values of $M_{\rm ej}$.

\begin{figure}
    \centering
     \vspace{-0.3cm}
    \includegraphics[width = 1\linewidth]{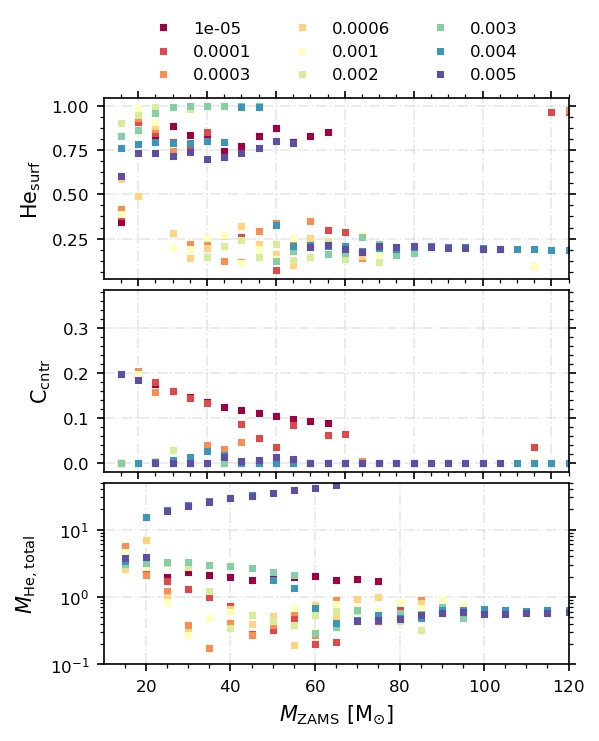}
    \vspace{-0.5cm}
    \caption{The top two sub-figures show the surface helium (He$_{\rm surf}$) and core carbon (C$_{\rm cntr}$) mass fraction contained in the star before the model halts.
    The bottom sub-figure shows the total helium mass $M_{\rm He, total}$ contained in the star at that instant. Some stars with $Z = 0.004, 0.005$ fail during core helium burning.}
    \label{fig: surf_he_cntr_c12}
\end{figure}

\subsubsection{Lack of He lines in the observed spectra of Ic-BL SNe}

Ic-BL SNe do not seem to contain traces of H or He in their spectra (e.g. \citealt{Modjaz_2016} and reference therein). Based on the analysis of a large data set of SNe Ib, SNe Ic in \cite{Liu_2016} and SNe Ic, SNe Ic-bl in \cite{Modjaz_2016}, the latter concluded that neither smearing of He lines nor insufficient mixing of $^{56}$Ni in the ejecta could explain the lack of He lines in the observed optical spectra.
On the other hand, our lowest $Z$ CH models in Fig. \ref{fig: surf_he_cntr_c12} still contain a significant fraction of helium on their surface. The same figure also shows the total helium mass $M_{\rm He, total}$ contained in the star. 
Though stars with $Z \lesssim 0.002$ contain a high helium surface mass fraction, their $M_{\rm He, total}$ content is less than (or would likely be less than) 1 M$_{\odot}$ by the time of core carbon depletion. This makes them a plausible candidate for Ic-BL if the remaining amount of helium could be hidden (e.g. accreted by the compact remnant if it is a BH as in the next section) or lost by the time of core collapse.
To compare, \cite{Modjaz_2016} set the limit on the total helium content of Ic-BL SN ejecta as $M_{\rm He, final} < 0.2$ M$_{\odot}$. 

Due to numerical reasons, some of our systems fail to evolve till core carbon depletion, but under similar conditions and $Z = 0.00034$, \cite{Aguilera-Dena_2018} (see their Table 2) found that the more massive CH stars ($M_{\rm ZAMS} \gtrsim 75$ M$_{\odot}$) would lose nearly all their surface helium by the time of core collapse.
Furthermore, by boosting the mixing efficiency by ten times, they find the stars to be completely devoid of surface helium even at lower masses.
It is unclear if such high mixing efficiencies are realized in nature. In the current study, stars with $Z \gtrsim 0.001$ seem unlikely to lose/burn their surface helium content by the time of core collapse which then disfavors them as a potential candidate for Ic-BL.

\begin{figure}
    \centering
    \vspace{-0.3cm}
    \includegraphics[width = 1\linewidth]{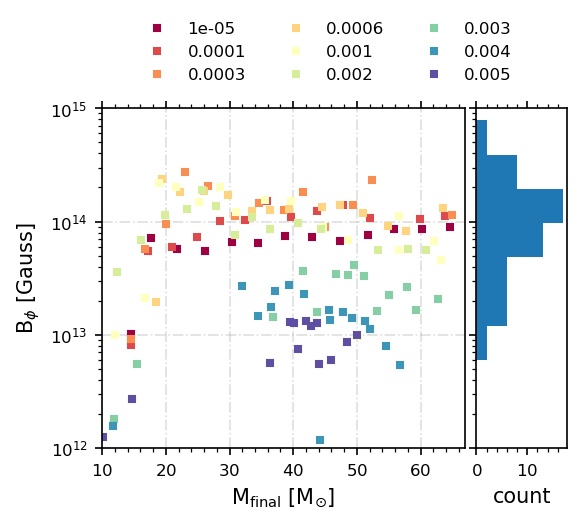}
    \vspace{-0.5cm}
    \caption{Mean azimuthal component of the NS's magnetic field formed from the collapse of the inner 1.6 M$_{\odot}$ baryonic mass of the CH star under the assumption of magnetic flux conservation (see, Eq. \ref{eq: magnetic_flux}). On the RHS are shown the inferred values of magnetic field for the SLSN sample analyzed  by \protect\cite{Nicholl_2017, Villar_2018}.}
    \label{fig: B_phi_vs_Mfinal}
\end{figure}

\subsubsection{Magnetic field strength of the proto neutron star}

Fig. \ref{fig: B_phi_vs_Mfinal} shows (and compares with observations) the mean azimuthal component of the magnetic field for a NS of 12 km radius formed from the inner 1.6 M$_{\odot}$ baryonic mass of the corresponding CH star. In doing so we assume that while the star contracts its magnetic flux $\Phi$ is conserved, i.e.

\begin{equation}
    \Phi_{\rm NS} = \int_{0}^{r(M=1.6 {\rm M}_{\odot})} B_{\phi} \pi r dr 
    \label{eq: magnetic_flux}
\end{equation}

\noindent However, it is expected that the NS's magnetic field would be further amplified by dynamo action during the proto-NS stage driven by convection \citep{Thompson_1993} or magnetorotational instability (e.g. see discussion in \citealt{Aguilera-Dena_2020}). This would further raise the magnetic field in Fig. \ref{fig: B_phi_vs_Mfinal} to values capable of producing Ic-BL ($\gtrsim 10^{15}$ G).



\subsection{Contribution towards GRBs and Ic-BL with a collapsar as the source} \label{subsec: Collapsar_implication}

Some studies have simultaneously observed GRBs and Ic-BL occurring from the same SN explosion (e.g. \citealt{Stanek_2003, Sollerman_2006}).
Though it can be expected for a GRB (under collapsar scenario) to occur without an accompanying Ic-BL SN as the latter requires a large energy reservoir, studies have also found Ic-BL seemingly originating from a central engine but without any accompanying GRB (e.g. \citealt{Soderberg_2010, Stevance_2017}).
This could simply be occurring due to the GRB jet being directed away from our line of sight but could also imply a more complex link between GRB/Ic-BL resulting from a rapidly rotating central source.
Nonetheless, here we will find that all CH stars contributing towards a Ic-BL (under the collapsar consideration) also produce an accompanying GRB.

Fig. (\ref{fig: Eiso_Ewind_vs_Mfinal}, \ref{fig: Eiso_Ewind_vs_MZAMS}) show the isotropic-equivalent energy $E_{\rm iso}$ produced and the energy carried by the winds $E_{\rm wind}$ during a long duration GRB under the collapsar scenario.
We again set the minimum energy for Ic-BL, $E_{\rm wind} = 10^{52}$ erg and adopt the value $E_{\rm iso} >10^{51}$ erg \citep{Perley_2016}.
To model the contribution of CH stars towards GRBs, we use the one-zone accretion disk evolution formalism as described in \cite{Kumar_2008} and summarised in \cite{Fuller_2022}.
We utilize the material made public by \cite{Fuller_2022} to perform our calculations for the result shown in Fig. (\ref{fig: Eiso_Ewind_vs_Mfinal}, \ref{fig: Eiso_Ewind_vs_MZAMS}) and here give a brief summary of the method employed to provide context.

\subsubsection{Estimating energy outflow from the accretion disk}

In the collapsar scenario, accretion of matter by a newly-born central black hole from a disk formed near the outer portion of the rapidly rotating stellar core \footnote{A stable disk is formed if the specific AM of the layer $j > j_{\rm ISCO}$ where $j_{\rm ISCO}$ is the specific AM of the innermost stable circular orbit for the BH.} powers the prompt emission of a relativistic jet.
After the initial formation of an accretion disk, \footnote{It is assumed that fallback of matter from the disk only occurs from the region making an angle > 30$^{\circ}$ with the rotational axis. The rest is assumed to be removed due to jet/winds.} it might undergo strong neutrino cooling due to the Urca process \citep{Gamow_1941, Qian_Woosley_1996} facilitating the transport of mass to smaller radii due to viscous drag.
In case of inefficient neutrino cooling, the system could enter the advection-dominated accretion flow (ADAF) regime where most of the matter will be blown away by strong disk wind.
The one-zoned accretion disk model (as outlined below) approximates  the combined actions of mass fallback, viscous accretion, and disk wind to determine the time evolution of the BH accretion disk.


Let the accretion disk rotate at Keplerian frequency with a characteristic disk radius $r_{\rm d}$. In the ADAF regime vigorous outflow continuously removes mass from the disk causing the accretion rate to decrease with radius. We assume that the accretion rate $ \dot M_{\rm acc} (r)$ varies as a power law with radius $r$ \citep{Blandford_1999} such that

\begin{equation}
    \dot M_{\rm acc} (r) = \frac{M_{\rm d}}{\tau_{\rm vis}} \left(\frac{r}{r_{\rm d}} \right)^{s},  \;\;\;\; (r_{\rm t} < r < r_{\rm d})
    \label{eq: accretion_power_law}
\end{equation}

\noindent where $M_{\rm d}$ is the mass of the disk, $M_{\rm d} / \tau_{\rm vis}$ is the mass accretion rate at $r_{\rm d}$, $\tau_{\rm vis} = \alpha^{-1} (r_{\mathrm{d}}^{3} / G M_{\mathrm{BH}})^{1/2}$ is the viscous timescale with $\alpha \in (0.01,0.1)$ being the dimensionless viscous parameter \citep{Shakura_Sunyaev_1973} and $s$ is a free parameter and could lie between 0.3-0.8 \citep{Yuan_Narayan_2014}. Here we take $\alpha = 0.03, s = 0.5$. Also $r_{\rm t}$ is the transitional radius below which the disk starts to be neutrino cooled resulting in efficient accretion (i.e. the power-law flattens) and is here calculated by setting (see, \citealt{Fuller_2022}) 

\begin{equation}
    \dot M_{\rm acc} (r_{\rm t}) = 10^{-2.5}  \frac{r_{\rm t}}{r_{\rm s}} \;\; {\rm M}_{\odot} {\rm yr}^{-1} 
\end{equation}

\noindent where $r_{\rm s}$ is the Schwarzschild radius. The differential rate of outflow of mass in winds is then 

\begin{equation}
    d \dot M_{\rm acc} = s \frac{M_{\rm d}}{\tau_{\rm vis}} \cdot \frac{dr}{r_{\rm d}^{s} r^{1-s}} 
    \label{eq: differential_mass_outflow}
\end{equation}

\noindent Assuming the gas particles in the wind escape to infinity gives $v_{\rm esc} = (GM_{\rm BH}/r)^{1/2}$ carrying a specific kinetic energy $GM_{\rm BH}/2r$. Multiplying this with Eq. \ref{eq: differential_mass_outflow} gives the differential rate of energy outflow and integrating that over $r$ from $r_{\rm t}$ to $r_{\rm d}$ gives the total rate of energy outflow in the wind as (e.g. \citealt{Kohri_2005})

\begin{equation}
L_{\mathrm{wind}} \simeq \frac{s}{2(1-s)} \frac{G M_{\mathrm{BH}}}{r_{\mathrm{d}}} \frac{M_{\mathrm{d}}}{\tau_{\mathrm{vis}}}\left[\left(r_{\mathrm{d}} / r_{\mathrm{t}}\right)^{1-s}-1\right],  \;\;\;\; s<1
\label{eq: wind_luminosity}
\end{equation} 

\noindent where $\frac{M_{\mathrm{d}}}{\tau_{\mathrm{vis}}} \equiv \dot M_{\rm acc}(r_{\rm {d}})$. On the other hand the BH mass gain rate is taken as $\dot M_{\rm BH} = \dot M_{\rm acc} (r_{\rm ISCO})$ where it is assumed that no mass is lost in disk wind once $r \leq r_{\rm ISCO}$, ISCO being the innermost stable circular orbit. The energy dissipation rate by this mass prior to accretion is approximated as 

\begin{equation}
L_{\mathrm{acc}} \simeq \eta_{\mathrm{NT}} \dot{M}_{\mathrm{BH}} c^{2}
\label{eq: accretion_luminosity}
\end{equation}

\noindent where $\eta_{\mathrm{NT}}=1-$ $\sqrt{1-r_{\mathrm{s}} / 3 r_{\text {isco }}}$ is the Novikov-Thorne efficiency parameter and is a measure of the energy released by a particle falling from infinity to ISCO \citep{Novikov_Thorne_1973}. Finally, integrating Eq. (\ref{eq: wind_luminosity}, \ref{eq: accretion_luminosity}) over the evolutionary timescale of the accretion disk gives the energy generated by the disk $E_{\rm acc}$ and the energy of the winds, $E_{\rm wind}$ \footnote{The lifespan of the disk is $\sim$ to the free-fall timescale of the progenitor star at core-collapse. For most cases we find this to be equal to 60-300 seconds.}. Only a fraction, $\eta E_{\rm acc}$ of the  energy generated by the accretion disk would be in the form of radiation that will escape in the jet.  There is also a fractional probability ($f \equiv d\Omega / 4 \pi$, where $d\Omega$ is the total solid angle spanned by the jet cones) that the jet aligns with the line of sight hence the total isotropic-equivalent energy released by the GRB is (e.g. \citealt{Ghirlanda_2004A, Bavera_2022}) 

\begin{equation}
    E_{\rm iso} = \frac{\eta E_{\rm acc}}{f}.
    \label{eq: E_iso}
\end{equation}

\noindent We choose $\eta = 0.01$ and $f = 0.01$ \citep{Guetta_Della_Valle_2007} such that $\eta / f = 1$. Changing this ratio would directly scale the value of $E_{\rm iso}$ in Fig. (\ref{fig: Eiso_Ewind_vs_Mfinal}, \ref{fig: Eiso_Ewind_vs_MZAMS}) by the same amount. We note that as discussed in Fig. \ref{fig: P_NS_E_rot_vs_Mfinal}, the stars with $Z  \geq 0.004$ fail during core helium burning and consequently should not feature in the GRB/Ic-BL region.

\subsubsection{Mass-metallicity range and formation rate of GRB}

\begin{figure}
    \centering
    \includegraphics[width = 1.02\linewidth]{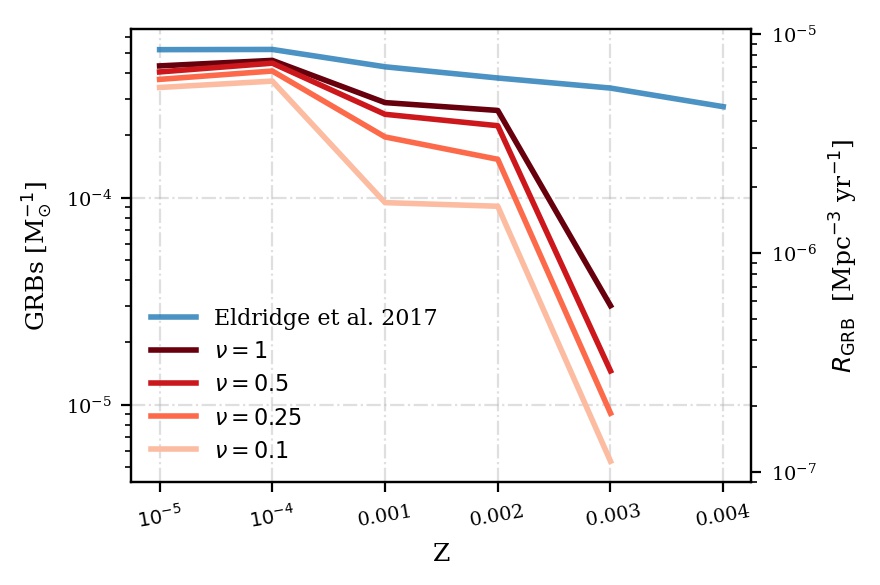}
    \vspace{-0.5cm}
    \caption{The number of potential GRBs per solar mass expected from the accretion CHE route within \textsc{bpass}.  On RHS of the y-axis is their local rate in Mpc$^{-3}$ yr$^{-1}$. Assuming a star formation rate of 0.015 M$_{\odot}$ Mpc$^{-3}$ yr$^{-1}$ \citep{Madau_Dickinson2014} occurring at a given value of $Z$ and negligible GRB delay time gives the local volumetric GRB rate  estimates $ \in  [\sim 5 \times 10^{-6}, \; \sim 10^{-7}]$ Mpc$^{-3}$ yr$^{-1}$.}
    \label{fig: GRB_vs_Z}
\end{figure}

Table \ref{tab:GRB_mass range} gives the allowed mass range for the accretors (near the end of the accretion phase) that can produce a GRB under the CH route. The cutoff $Z$ (depending on the nature of winds) is near $Z \sim 0.004$. Many of these GRB are also capable of producing $E_{\rm wind}$ that could power a Ic-BL SN.
Fig. \ref{fig: GRB_vs_Z} shows the rate of GRB formation per solar mass of star formation in \textsc{bpass}. This gives a local volumetric GRB rate density (as a function of $Z$) $ \in  [\sim 5 \times 10^{-6}, \; \sim 10^{-7}]$ Mpc$^{-3}$ yr$^{-1}$ assuming a star formation rate of 0.015 M$_{\odot}$ Mpc$^{-3}$ yr$^{-1}$ \citep{Madau_Dickinson2014} and negligible GRB delay time.

\begin{figure}
    \centering
    \vspace{-0.5cm}
    \includegraphics[width = 1\linewidth]{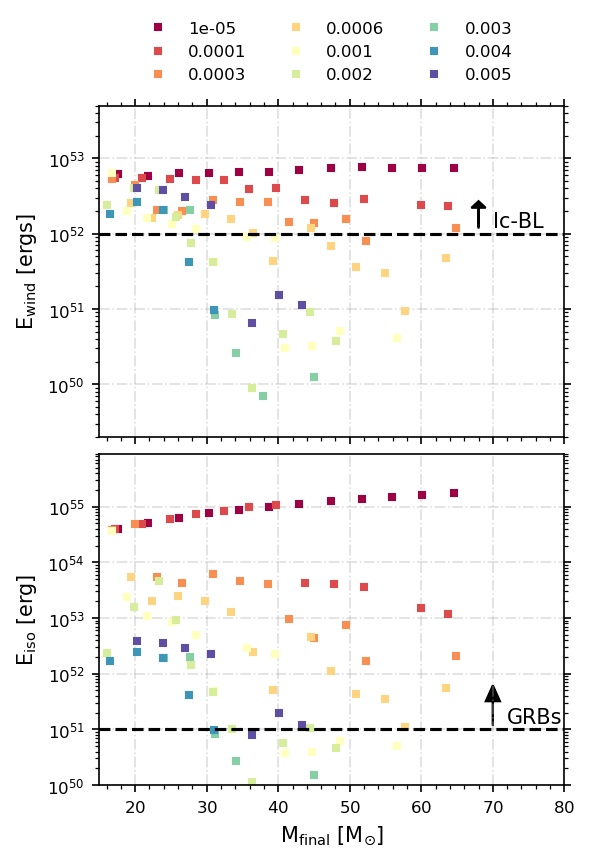}
    \vspace{-0.5cm}
    \caption{The isotropic-equivalent energy $E_{\rm iso}$ produced and the energy carried by the winds $E_{\rm wind}$ during a long duration GRB under the collapsar scenario for the CH stars. We set the minimum energy for Ic-BL, $E_{\rm wind} = 10^{52}$ erg and adopt the value $E_{\rm iso} >10^{51}$ erg \citep{Perley_2016}. The top (bottom) sub-figure is obtained by integrating Eq. \ref{eq: wind_luminosity} (Eq. \ref{eq: accretion_luminosity} with $\eta, f = 0.01$) over the evolutionary time of the accretion disk for the corresponding CH star.}
    \label{fig: Eiso_Ewind_vs_Mfinal}
\end{figure}

The rates at larger value of $Z$ are generally in agreement with the expected value of $> 10^{-7}$ Mpc$^{-3}$ yr$^{-1}$ (e.g. \citealt{Guetta_Della_Valle_2007}) keeping in mind that only a fractional amount of star formation would occur at low values of $Z$ in the local Universe. These rates would be further reduced by around two orders of magnitude if the source alignment with respect to the observer is taken into account (for beaming factor $f = 0.01$).
We note that for these accretors the mass range is set by the mass of the accretor near the end of the accretion phase (as this decides if it could become CH, Eq. \ref{eq: Packet_modified}) and not $M_{\rm ZAMS}$. Depending on the value of the angular momentum accretion efficient $\nu$ and the initial metallicity, the allowed accretor's $M_{\rm ZAMS}$ range can be approximated as $M_{\rm ZAMS} = M_{\rm min/max} /(1 + \Delta M / M_{\rm ZAMS})$ where $\Delta M / M_{\rm ZAMS}$ is derived from Eq. \ref{eq: Packet_modified}. The minimum (maximum) $M_{\rm ZAMS}$ should span 16-26 M$_{\odot}$ (48-78 M$_{\odot}$) for $r_{\rm g}^{2} = 0.05$ for the various values of $Z$, e.g. see Fig. \ref{fig: Old_vs_new_accrt_CHE}. Lastly, from Fig. \ref{fig: Eiso_Ewind_vs_Mfinal} we can see that many stars capable of producing a GRB are also a progenitor of Ic-BL. Hence the rate of Ic-BL from the CH channel should roughly follow the rate of GRB under the collapsar scenario.




\begin{table}
	\centering
    \caption{Allowed mass range for CH accretors that could potentially produce a GRB (cf. Fig. \ref{fig: Eiso_Ewind_vs_Mfinal}, \ref{fig: Eiso_Ewind_vs_MZAMS}). Where some models fail to evolve  near the lower or upper end of the mass range, we adopt the mass cutoff values from the nearest larger $Z$. We note that for these accretors the mass range is set by the mass of the accretor near the end of the accretion phase $M^{\rm post, acc}$ (as this decides if it could become CH in Eq. \ref{eq: Packet_modified}) and not $M_{\rm ZAMS}$.}
    \label{tab:GRB_mass range}
    \begin{tabular}{rrr}
    \hline
     $Z$    &     $M_{\rm min}^{\rm post, acc}$ [M$_{\odot}$] &  $M_{\rm max}^{\rm post, acc}$ [M$_{\odot}$] \\
    \hline \hline
     0.00001 &      20 &      75 \\
     0.0001 &      20 &      80 \\
     0.0003 &      20 &      85 \\
     0.0006 &      20 &      85 \\
     0.001 &      20 &      55 \\
     0.002 &      25 &      55 \\
     0.003 &      30 &      35 \\
    \hline
    \end{tabular}
\end{table}


\section{Caveats and Uncertainties}   \label{sec: Discussion}

Apart from the typical uncertainties involved in single (e.g. \citealt{Buldgen_2019}) and binary stellar evolution (e.g. \citealt{Ivonova2013}), below we mention the caveats and uncertainties specifically applicable to the current work.


\subsection{General caveats and uncertainties}
\begin{enumerate}
    
    \item The results shown here are strongly dependent on the assumption made in section \ref{sec: Accretion CHE versus CHE in isolated star}. In short, we assume that the overall evolution of a rapidly-rotating single star leading to CHE can also be realized in the secondary component of a binary system under sufficient accretion (e.g. \citealt{Cantiello2007}) as evaluated using Eq. \ref{eq: Packet_modified}. 
    
    
    \item Eq. \ref{eq: threshold_angular_velocity_single_star} was derived for stars experiencing CHE from the onset of ZAMS. While it is likely that the accretor star would have developed some level of chemical gradient before it begins to accrete AM. Therefore employing Eq. \ref{eq: threshold_angular_velocity_single_star} to determine the amount of mass required to be accreted for CHE in Eq. \ref{eq: Packet_modified} implicitly assumes that the inhibiting effect of the accretor's chemical gradient on chemical mixing in minimal.
    Moreover, if accretion of the material occurs at a later stage of the secondary’s evolution such that its convective core does not have enough time to adapt to the increased mass then rejuvenation of the star due to mixing could be altogether avoided (e.g. \citealt{Braun_Langer1995}). The radius of gyration $r_{\rm g}$ of such an evolved star would be relatively smaller and hence Eq. \ref{eq: Packet_modified} would lead to a lower value of $\Delta M / M$. This would bias these evolved systems towards CHE under our current setting. Currently we set a rough cutoff value of $r_{\rm g}^{2} > 0.025$ for an accretor to potentially experience CHE. This is a crude approximation to filter out the highly evolved systems and more analysis would be required to place a better constraint.

    \item Uncertainties in the exact nature of stellar wind mass loss and the coupling of stellar winds to the orbit (hence modifying the orbit) could affect the outcome of our calculation for the case of tidal CHE. We assume the wind to be fast and isotropic and only taking away the orbital AM equivalent to the specific AM of the mass-losing star which might not necessarily be the case. Furthermore, the stellar winds directly dictate the amount of AM that could be retained by an accretion induced CH star. As a result, stronger winds could substantially change the results shown in this paper.
     
     \item The results shown in section \ref{subsec: Magnetar_implication} and \ref{subsec: Collapsar_implication} are highly dependent on the assumption that the pre core collapse AM content of the star is well approximated by its AM content at core carbon depletion. While the evolutionary trend of these \textsc{mesa} models near core carbon depletion justifies our assumption (e.g. Fig. \ref{fig: ST_vs_efficient_AM}) a proper 3D treatment of AM transport near core-collapse is desirable. Recently, \cite{Fields_2022} studied the 3D hydrodynamic evolution of a 16 M$_{\odot}$ rapidly rotating star during the final 10 minutes prior to iron core collapse. They found a reasonable match between the iron core's AM structure of their model and the one from \textsc{mesa} with the resulting NS's period being within 5 per cent agreement with a caveat that difference in the Si/O convective structure in the progenitors could lead to significant deviation in the AM profiles of the 1D and 3D models. Nonetheless,  consensus on the efficiency of rotational induced mixing (e.g. \citealt{Hunter_2008_surface_anomalies, Aguilera-Dena_2018}) and the effectiveness of 1D stellar evolution code in determining the stellar evolution for these rapidly rotating stars remains ambiguous. 
     
     \item Here we adopt a standard treatment for the efficiency of chemical mixing (e.g. \citealt{Yoon_2005}) and angular momentum transport (e.g. \citealt{Zahn_1992}) in evaluating Eq. \ref{eq: threshold_angular_velocity_single_star}. We note that previously \cite{Aguilera-Dena_2018, Aguilera-Dena_2020} have employed a larger efficiency of mixing though it is unclear if such high mixing efficiencies are realized in nature.
    \end{enumerate}
     
     \subsection{Caveats/Uncertainties due to employing  pre-computed \textsc{bpass} models for conducting population study}
     
     \begin{enumerate}

     \item In \textsc{bpass}, mass accretion on a star is limited by its thermal time scale. As a result, some secondaries accrete more mass than is required to spin them up to their breakup velocity (as accretion does not cease once the star approaches its breakup). 
     Since \textsc{bpass} models are pre-processed, it is time consuming to further change the models to suit this study.
     Nevertheless, approximating the disk boundary layer as a polytrope and employing the $\alpha$-prescription \citep{Shakura_Sunyaev_1973} to model viscosity,  \cite{Paczynski1991} and  \cite{Popham_Narayan_1991} argued that the presence of viscous torques due to differential rotation can efficiently transport just as much AM outwards from the stellar surface as the accreting matter brings.
     In such a case the star can still keep accreting matter from the accretion disk without gaining much AM. Under this scenario, the secondary accretion in \textsc{bpass} can be validated.
 
    \item In \textsc{bpass}, initial stellar rotation is enforced using the analytical fits of \cite{Hurley_2000} to the data of \cite{Lang_1992}. For simplicity here we assume that all stars prior to accretion have zero rotational velocity, $\omega_{\rm in}$. This is an underestimate and stars with already finite $\omega_{\rm in}$ at the start of the accretion phase would require relatively less amount of accretion to spin up leading to an underestimate in the number of accretion CH stars.
    
     \item Tidal effects on the accretor are only considered once the episode of mass accretion ends. This might unfairly prevent those CH secondaries from experiencing tidal spin-down that undergo mass accretion on a nuclear timescale. Additionally, while checking for tidal spin-down of the secondary star, we do not let the separation between the binary to evolve due to mass loss. This could overestimate the efficiency of tidal spin down, consequently lowering the frequency of accretion CHE stars in our study.

     \item \textsc{bpass} does not yet account for systems with initial mass-ratio $q_{i} = 1$ or binaries with initial orbital period less than 1 day. Consequently, we might be underproducing equal mass tidal CH systems or tidal CH systems in very small initial periods. Nevertheless, we expect these systems to be in roughly similar proportion to those represented with blue color in Fig. \ref{fig: CHE_percent}. This is because otherwise, as per our calculation, the subsequently formed binary black holes resulting from the tidal CHE stars may produce a detectable black hole merger rate larger than that observed by the LIGO-VIRGO-KAGRA network.

    
\end{enumerate}

    
\section{Summary \& Conclusions} \label{sec: conclusion}

In this paper, we have discussed two potential pathways leading to rapidly rotating stars in binary systems. Following standard treatment for rotational induced mixing using the 1D stellar evolution \textsc{mesa}, we find that for sufficiently rapid rotation, stars could subsequently experience CHE. Consequently, this would lead to the formation of homogeneous helium stars and depending on the initial mass, metallicity and the formation pathway, they could even sustain a substantial fraction of their AM until core collapse. 

We calculate the threshold minimum initial angular velocity required by a single star as a function of its $Z$ and $M_{\rm ZAMS}$ to undergo CHE (Fig. \ref{fig: Threhold_omega_Dutch_factor_variable}, Eq. \ref{eq: threshold_angular_velocity_single_star}). We generalize the analytical relation in \cite{Packet1981} to now give the threshold mass accretion to spin up a star to any desired (below critical) angular velocity (Eq. \ref{eq: Packet_modified}). We then use the relations in Eq. \ref{eq: threshold_angular_velocity_single_star} in conjunction with Eq. \ref{eq: Packet_modified} to calculate the threshold mass accretion required by a star in binary to experience CHE. Next, we implement this into  \textsc{bpass} to study its effect on a synthetic stellar population. We also incorporate tidal CHE in \textsc{bpass} following the work of \cite{Riley_2021}. Our core findings are as follows:

\begin{enumerate}
    \item Accretion induced CHE is the dominant means of producing homogeneous stars when compared to tidal CHE.
    
    \item In contrast to tidal CHE, accretion induced CHE can produce stars that retain a substantial fraction of their acquired AM. As a result, these stars can be potential  progenitors of exotic transients like SLSNe, Ic-BL and long GRBs. To determine the suitability of a CH star in generating such energetic transients, we calculate the value of $P_{\rm NS}, E_{\rm rot}$ and $B_{\phi}$ of the resulting proto-NS and $E_{\rm iso}$ and $E_{\rm wind}$ of the resulting BH in section \ref{subsec: Magnetar_implication} and \ref{subsec: Collapsar_implication}.
    
    \item For certain systems a single progenitor could produce both SLSNe and Ic-BL (GRB and Ic-BL) under the magnetar (collapsar) scenario. 
    
    \item CH stars have high surface temperatures and as such strongly emit at $\lambda < 912$\AA. The emission rate predicted here (Fig. \ref{fig: ionising_photons_count}) approximately matches the earlier findings (\citealt{Bpass2017}) at low values of $Z$. Disagreement occurs at $Z \gtrsim 0.001$ where we find that the ionizing photon density could decrease by 15-65 per cent depending on the value of $\nu$ and the metallicity.

\end{enumerate}

  If accretion is the primary driver of CH stars, then population studies that only consider tidal CHE may be underestimating the importance of this evolutionary channel. While observational data remains inconclusive on the occurrence of CHE and the role of rotation in their formation, such systems if realized could potentially be one of the primary drivers of transients like LGRBs, Ic-BL, SLSN-I and also play a substantial role in cosmic reionization.

\section*{Acknowledgements}

We are grateful to Georges Meynet for commenting on the manuscript and providing valuable feedback. Sohan Ghodla acknowledges support from the University of Auckland. J. J. Eldridge and Héloïse F. Stevance acknowledge the support of the Marsden Fund Council managed through the Royal Society of New Zealand Te Apārangi. Elizabeth R. Stanway received support from the United Kingdom Science and Technology Facilities Council (STFC) grant ST/T000406/1. This project utilised NeSI high performance computing facilities, python, matplotlib \citep{Matplotlib_2007}, numpy \citep{Numpy_2020}) and pandas \citep{Pandas_2020}. We thank \cite{Fuller_2022} for making their code public.


\section*{Data Availability}


The data underlying this article will be shared on reasonable request to the corresponding author. The \textsc{mesa} models and the python notebook used to compute some of the figures are available at \href{https://github.com/SohanGhodla/CHE-and-its-EM-implications}{at this Github address.}




\bibliographystyle{mnras}
\bibliography{refs} 




\appendix
\phantom{xxxx}

\section{The minimum threshold accretion for chemically homogeneous evolution} \label{Sec1: Appendix}

Let $M,R, r_{g}, I, L, \omega_{\rm in}$ be the mass, radius, radius of gyration, moment of inertia, luminosity and the angular velocity of the secondary star at the initiation of accretion. Then the moment of inertia can be written as $I = r_{\rm g}^{2} MR^{2}$.
The effect of radiation pressure on the infalling matter at the stellar surface can be modeled using the Eddington factor, $\Gamma = L/L_{\rm edd}$ where $L_{\rm edd} = \frac{4 \pi G M c}{\kappa}$ is the Eddington Luminosity. Hence, the critical angular velocity of the star is
\begin{equation}
    \omega_{\text {cr}}=\sqrt{\frac{G M (1 - \Gamma)}{R^{3}}}
    \label{eq: omega_cr}
\end{equation} 



It is assumed (as in \citealt{Packet1981}) that the matter is being accreted from the inner ring of the disk formed around the star's equator where the matter is moving in approximately Keplerian orbit with negligible viscosity. Therefore the angular velocity of the matter in the ring above the equatorial surface is $\omega_{\text {ring }} = \omega_{\text {cr}}$.  Moreover the specific AM carried by the mass element about to be accreted is
\begin{equation}
     j = \frac{I \omega_{\rm ring}}{m} = R^{2} \sqrt{\frac{G M (1 - \Gamma)}{R^{3}}}  = \sqrt{GMR(1 - \Gamma)}
\end{equation} 



We assume $R, \Gamma$ to be constant in the following calculation with magnitude equal to their values at the beginning of mass transfer. We will later find that for systems at low $Z$ only a small amount of $\Delta M$ is needed for the star to experience CHE and hence $\Delta R$ would be minimal especially if $R$ is already large. Additionally, if we approximate the luminosity as a power-law relation $L \sim M^{\alpha}$ and take $\alpha = 1$ for very massive stars i.e. $M$ > 55 M$_{\odot}$, then at least for these stars $\Gamma$ also approaches a near constant value. Now $\Delta J$ added by accretion of mass $\Delta M$ from the ring is

\begin{equation}
    \begin{aligned}
         \Delta J & = \nu \int_{M}^{M+\Delta M}  j(m) d m  = \nu \int_{M}^{M+\Delta M} \sqrt{GmR(1 - \Gamma)} \; dm \\ & =\frac{2 \nu}{3} \sqrt{GR(1 - \Gamma)} \left[(M+\Delta M)^{3 / 2}-M^{3 / 2}\right]  
    \end{aligned}
    \label{eq: Appendix_A4}
\end{equation}




\noindent where $\nu$ is a free parameter that encodes the accretion efficiency of AM by the star from the Keplerian disk \footnote{The inwards transport of the accreted AM from the surface could be carried by the Spruit-Tayler dynamo induced magnetic fields as well as the rotationally induced instabilities and meridional currents.}. One can instead assume full accretion efficiency of AM but now happening from the inner edge of a sub-Keplerian disk rotating at angular frequency $\nu \omega_{\rm cr}$ and still get the same result. Eq. \ref{eq: Appendix_A4} can be further simplified as

\begin{equation}
\begin{aligned}
\Delta J & = \frac{2\nu }{3} \sqrt{GR(1 - \Gamma)} M^{3 / 2} \left[\frac{3}{2} \frac{\Delta M}{M}+\frac{3}{4} \left(\frac{\Delta M}{M}\right)^{2} +O(3) \right] \\
& \approx \nu \sqrt{GR(1 - \Gamma)} \Delta M\left(\sqrt{M}+\frac{\Delta M}{4 \sqrt{M}}\right)
\end{aligned}
\end{equation} 
where we have Taylor expanded the term in the square bracket and neglected terms of order three and higher. This is a fair approximation for small values of $\Delta M / M$ which is typically the case here (in Fig. \ref{fig: Acc_CHE} $\Delta M / M \lesssim 0.9$). Now if we let $J^{\prime}$ be the new AM of the secondary, then
\begin{equation}
J^{\prime} = J + \Delta J = I \omega_{\rm in} + \Delta J = (I+\Delta I) \omega_{\text {CHE }}
\end{equation}

\noindent where we accrete at least the threshold minimum mass required for CHE but $\omega_{\rm CHE}$ could easily be replaced by any desired value of angular velocity. This implies

\begin{equation}
\begin{aligned}
    & \left(M(r_{\rm g} R)^{2}+\Delta M\left(r_{\rm g} R\right)^{2}\right) \omega_{\text {CHE }} \\ 
  & = M\left(r_{\rm g} R\right)^{2} \omega_{\rm in} + \nu \sqrt{GR(1 - \Gamma)} \Delta M\left(\sqrt{M}+\frac{\Delta M}{4\sqrt{M}}\right) 
\end{aligned}
\end{equation}

\noindent Above $r_{\rm g}$ and $R$ were treated as constants in calculating $\Delta I$. Though their values would change during the course of accretion, as \cite{Packet1981} argues, the effect of matter accretion is such that while $R$ grows, $r_{\rm g}$ decreases. Consequently, they to a certain degree balance each other's effect. Now using Eq. \ref{eq: omega_cr} and rearranging the above equation gives



\begin{equation} \label{eq:Appendix}
\frac{\Delta M}{M} = \pm \sqrt{\frac{1}{4}\left(r_{\rm g}^{2} \frac{\omega_{\rm CHE}}{\nu \omega_{\rm cr}}+1\right)^{2}-r_{g}^{2} \frac{\omega_{\rm in}}{\nu \omega_{\rm cr}}}-\left(\frac{1}{2}-\frac{r_{\rm g}^{2} \omega_{\rm CHE}}{2 \nu \omega_{\rm cr}}\right)
\end{equation}

\noindent where the negative solution leads to negative values of accreted mass and hence is neglected. Substituting $\omega_{\rm CHE} = \omega_{\rm cr}$ and $\nu = 1$  in Eq. \ref{eq:Appendix} reduces this equation to the one derived in \cite{Packet1981}. Lastly, the $\Gamma$ factor is already taken into account while calculating $\omega_{\rm CHE}/\omega_{\rm cr}$ expression in Eq. \ref{eq: threshold_angular_velocity_single_star}.

\section{Fits for Dutch scaling = 0.5}   \label{sec: Dutch_scaling_05}



The exponents in Eq. \ref{eq: threshold_angular_velocity_single_star} for the \cite{Nugis_Lamers_2000} wind scheme with the Dutch factor = 0.5 are as follows:

\begin{equation}
\begin{array}{l}
\alpha(Z) = -9.382 \times 10^{6} Z^{3} -152.5 Z^{2} + 518.5 Z -11.67 \\
\beta(Z) = -1.546 \times 10^{7} Z^{3} +6.55 \times 10^{4} Z^{2} + 118.4 Z -5.997 \\
\gamma(Z) = -1.528 \times 10^{6} Z^{3} +10.57 Z^{2} + 106.8 Z-0.561
\end{array}
\label{eq: Threshold_CHE_Dutch_05}
\end{equation}

\begin{figure}
    \centering
    \vspace{-0.5cm}
    \includegraphics[width = 1\linewidth]{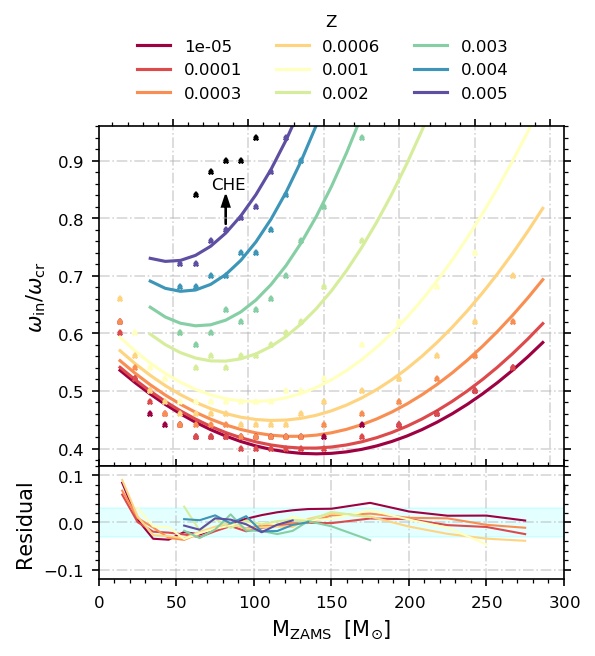}
    \vspace{-0.5cm}
    \caption{Same as Fig. \ref{fig: Threhold_omega_Dutch_factor_variable} but now with the Dutch scaling factor = 0.5. The black dots represent $Z = 0.006$ that could not be fitted.}
    \label{fig: Threhold_omega_Dutch_factor_05}
\end{figure}

\noindent The corresponding $\omega_{\rm CHE} / \omega_{\rm cr}$ relation is plotted in Fig. \ref{fig: Threhold_omega_Dutch_factor_05}. We note that for the above winds we do find a few CHE systems at $Z = 0.006$, but we have not included them in Eq. \ref{eq: Threshold_CHE_Dutch_05} since they cannot be fitted to a cubic relation.

\section{Additional figures}

\begin{figure}
    \centering
    \includegraphics[width = 0.95\linewidth]{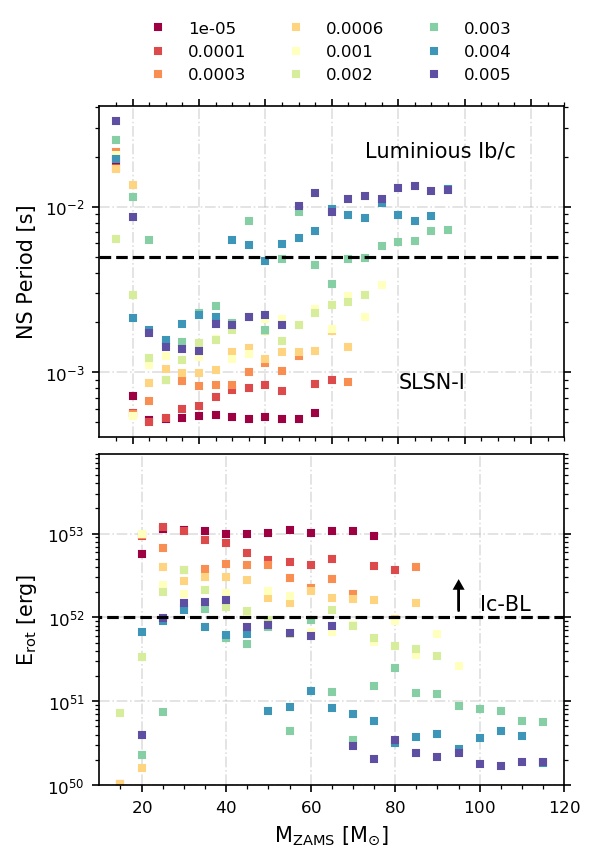}
    \vspace{-0.3cm}
    \caption{Same as Fig. \ref{fig: P_NS_E_rot_vs_Mfinal} but now showing $M_{\rm ZAMS}$ on the x-axis.}
    \label{fig: P_NS_Erot_MZAMS}
\end{figure}

\begin{figure}
    \centering
    \includegraphics[width = 0.95\linewidth]{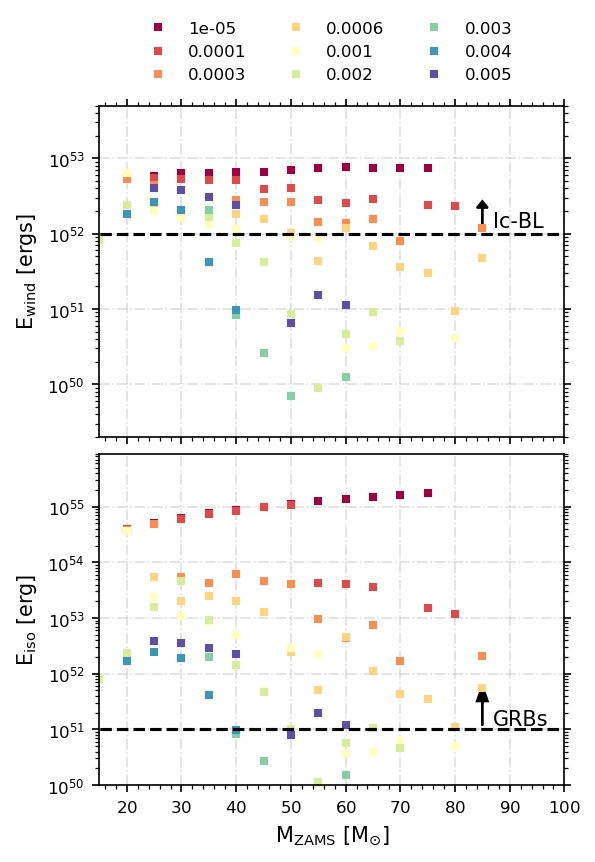}
    \vspace{-0.3cm}
    \caption{Same as Fig. \ref{fig: Eiso_Ewind_vs_Mfinal} but now showing $M_{\rm ZAMS}$ on the x-axis.}
    \label{fig: Eiso_Ewind_vs_MZAMS}
\end{figure}






\bsp	
\label{lastpage}
\end{document}